\documentclass[12pt]{amsart}

\usepackage{latexsym, amsmath, amssymb,amsthm, natbib,verbatim, xy, microtype}
\usepackage{color,soul}
\xyoption{all}

\usepackage{tgpagella}
\usepackage{mathpazo}
\usepackage{enumerate}
\usepackage{tcolorbox}
\tcbuselibrary{breakable,skins}

\usepackage{xcolor}
\colorlet{drkblue}{blue!6.8!black}
\usepackage[bookmarks=false]{hyperref}
\hypersetup{
 pdfborder = {0 0 0},
 colorlinks=true,
 citecolor=drkblue,
 linkcolor=drkblue,
 urlcolor=drkblue,
pdfstartview={XYZ null null 0.80}
}

\usepackage{tikz}

\usepackage[titletoc,title]{appendix}

\usepackage{etoolbox}
\patchcmd{\section}{\scshape}{\bfseries}{}{}
\makeatletter
\renewcommand{\@secnumfont}{\bfseries}
\makeatother

\usepackage[left=1.25in,top=1.25in,right=1.25in,bottom=1.25in]{geometry}
\usepackage{setspace,color}
\usepackage{graphicx,float}
\usepackage{enumerate}
\usepackage[multiple]{footmisc}
\usepackage[leqno]{amsmath}

\theoremstyle{plain}
\newtheorem{theorem}{Theorem}
\newtheorem{corollary}{Corollary}
\newtheorem{lemma}{Lemma}
\newtheorem{proposition}{Proposition}
\theoremstyle{definition}
\newtheorem{definition}{Definition}

\theoremstyle{remark}

\def\A{\mathcal{A}}
    
   \def\C{\mathcal{C}}

\usepackage{multirow,array}

\def\os{\emptyset}

\newcommand{\abs}[1]{\left| #1 \right|}

\begin{document}

\title[Market Design with Deferred Acceptance]{Market Design with Deferred Acceptance:\\ A Recipe for Policymaking}

\author[Do\u{g}an, Imamura, and Yenmez]{Battal Do\u{g}an \and Kenzo Imamura  \and M. Bumin Yenmez}

\thanks{\emph{Keywords}: Choice rules, punctual choice axioms, matching rules, deferred acceptance.\\
Do\u{g}an is affiliated with the Department of Economics, University  of  Bristol,
Bristol, UK; Imamura is with the Department of Economics, the University of Tokyo, Tokyo, Japan; Yenmez is with the Department of Economics, Boston College, Chestnut Hill, MA, USA. Emails: \texttt{battal.dogan@bristol.ac.uk}, \texttt{imamurak@e.u-tokyo.ac.jp},
\texttt{bumin.yenmez@bc.edu}. We are grateful to Fuhito Kojima and participants at the Saarland Workshop in Economic Theory for helpful comments.}

\begin{abstract}
We introduce a method to derive from a characterization of institutional choice rules (or priority rules), a characterization of the Gale-Shapley deferred-acceptance (DA) matching rule based on these choice rules. We apply our method to school choice in Chile, where we design choice rules for schools that are uniquely compatible with the School Inclusion Law and derive a set of matching properties, compatible with
the law, that characterizes the DA rule based on the designed choice rules. Our method provides a recipe for establishing such results and can help policymakers decide on which allocation rule to use in practice.
\end{abstract}



\maketitle

\section{Introduction}
Market design has been exceptionally successful in bridging the gap between
economic theory and real-life markets. Many markets have
been organized by economists so that desirable outcomes can be achieved.
The leading examples include the entry-level labor market for new physicians
in the US, school choice around the world, keyword auctions on internet search engines, spectrum auctions in many countries, and kidney exchange.\footnote{See, among others, \cite{roth_jama} for the residency match, \cite*{abdulka05b} and \cite*{abdul05} for school choice, \cite{ostrovsly_schwarz11} for keyword auctions, \cite{binmore_klemperer} for spectrum auctions, and \cite{keaerpp2005} for kidney exchange.}

Often, policymakers need to be persuaded to implement allocation rules that
produce desirable outcomes. To this end, it is crucial to have a set of properties that
pins down an allocation rule.
If policymakers view these properties as desirable, then the unique rule
satisfying the properties can be implemented. In the context of school choice,
fairness considerations led to the implementation of the deferred-acceptance (DA)
rule \citep{gale62},\footnote{The agent-proposing deferred-acceptance rule is the unique rule that is \emph{stable} and \emph{strategy-proof} \citep{albar94}. Both
these properties are viewed as fairness notions in the context of school choice.}
which is now used throughout the world in different contexts
such as assigning doctors to residency positions and assigning students to schools
or colleges.\footnote{See \cite{abdul_andersson} for a review of the school choice literature.}

To use the DA rule, however, policymakers first need to determine
\textit{institutional choice rules} that are the algorithmic proxies for
institutions in the course of the DA algorithm. These choice rules specify how institutions select from any set of applicants.\footnote{An institutional choice rule may reflect the preferences of the corresponding institution, objectives of a social planner, or both. A distinguishing feature of these choice rules from classical choice functions is their \textit{combinatorial} nature \citep{echenique_counting}:  institutions may combine several elements into a bundle to determine the choice, such as a school admitting many students from a pool of applicants. See \cite{alva_dogan} for a textbook chapter on modeling choice behavior and designing institutional choice rules for market design.} Sometimes they can be specified
simply by a single ranking over the applicants and a target capacity and sometimes they can be more complex, for example, when institutions have distributional objectives. The need for a twofold design (priority design and allocation rule design) brings about a need for a twofold justification: a characterization of the institutional choice rules and a characterization of the DA rule based on these choice rules.\footnote{Choice rules
and the DA rule based on some choice rules are often
considered separately, although they are intimately related. For example,
\cite{albar94}, \cite{kojman10}, and \cite{EHLERS2016128} provide characterizations of the DA rule while \cite{echyen12} provide
characterizations of choice rules for institutions with diversity considerations. Recently, a few papers have provided characterizations of some choice rules in specific settings together with a
characterization of the DA rule based on these choice rules \citep{doganJPET,aytur,cadet,sonmez_yenmez_india2}.}

In this paper, we introduce a method to derive from a characterization of institutional choice rules, a characterization of the DA rule based on these choice rules. \emph{Punctual} choice axioms turn out to be central to our analysis. A choice axiom is punctual if, on each choice problem, it imposes a requirement that can be checked without referring to other choice problems.\footnote{We borrow the \emph{punctual axiom} terminology from \cite{thomson_2012}, who
	makes several taxonomic suggestions to classify axioms for resource
	allocation problems.} We introduce a systematic way to extend any given punctual choice axiom to a matching axiom. We show that, if each institution is endowed with a choice rule that is characterized by a set of punctual choice axioms and, furthermore, satisfies \textit{path independence} and \textit{size monotonicity},\footnote{In other words, \textit{path independence} and \textit{size monotonicity} (defined in Section \ref{sec:model}) are implied by the set of punctual choice axioms. \textit{Path independence} and \textit{size monotonicity} are not punctual axioms as we discuss in Section \ref{sec:punctual}, while they are typically common to all choice rules in the literature in the context of market design with deferred acceptance because they ensure that the DA rule is strategy-proof and \textit{stable} \citep{hatfi04}.} then the DA rule based on these choice rules is characterized by \textit{individual rationality}, \textit{strategy-proofness}, and the extensions of the punctual choice axioms (Theorem \ref{thm:main}).

On the practical side, our method will come in handy especially when the institutional choice rules reflect complex objectives and are already difficult to justify. Note that it is easier to derive a characterization of choice rules rather than the DA rule based on these choice rules since a choice problem for an institution is essentially a simple matching problem that
includes only this institution. Moreover, on the conceptual side, our study sheds light on which properties of institutional choice rules are reflected in the market outcome when the DA rule is in use, and in particular, we establish that punctual choice axioms are always reflected in the market outcome.

Our method is widely applicable and can already be used to obtain characterizations of the DA rule based on well-known choice rules from the literature.\footnote{For example, choice rules with reserves introduced in \cite{hayeyi13}, which have been applied in practical applications, satisfies \textit{path independence} and \textit{size monotonicity} and, furthermore, is characterized by a set of punctual choice axioms \citep{imamura2020meritocracy}. Applying our method yields a characterization of the DA rule based on choice rules with reserves. \cite{echyen12} also provide a characterization of choice rules with reserves but their characterization
includes non-punctual axioms. See also
\cite{abdulkadirouglu2021priority}.} We consider several applications. In our main application, we study school choice in Chile where we design a novel choice rule that is uniquely compatible with the objectives of the policymakers and, using Theorem \ref{thm:main}, derive a characterization of the DA rule based on this choice rule.

In Chile, K-12 students are assigned to schools in a centralized clearinghouse nationwide \citep{correa2022}.\footnote{The clearinghouse in Chile was designed and implemented by the
authors of \cite{correa2022}.} In May 2015, the Chilean government passed the \textit{School Inclusion Law}.
According to the law, at each school, returning students have enrollment guarantees.
Furthermore, students are prioritized depending on some criteria such as whether
they have a sibling attending the school or a parent working at the school.
In addition, there are minimum-guarantee reservations for students with disabilities, economically
disadvantaged students, and students with high academic performance.\footnote{In the market-design literature, reserves were introduced by \cite{hayeyi13}.} \cite{correa2022} design a choice rule in which each seat has a
priority ranking and students are chosen one by one for each seat according
to an ordering of seats.\footnote{Even though it is not in the law, the design of
\cite{correa2022} assigns siblings to the same school. We ignore siblings in our analysis
because it is not required by the law that siblings are assigned to the same school and
it creates complementarities between students. It is well known that DA does not work well under complementarities \citep{hatfi04}.} Instead, we provide a choice rule that has enrollment guarantees and
provides reservations using a single priority ranking. We argue that our
choice rule is the unique choice rule
compatible with desirable properties in the spirit of the law (Proposition \ref{proposition_GE}). This result, however, only provides justification
for what priorities should be used at each school, but does not justify why
the DA rule should be implemented.
By applying Theorem \ref{thm:main}, we  derive a set of matching properties compatible with the law and show that the DA rule based on the designed choice rule uniquely satisfies these properties.

In the next section, we provide our model. In Section \ref{sec:punext}, we
formally introduce punctual choice axioms and their extensions to
matchings. We present the main result in Section \ref{sec:DA} and
applications in Section \ref{sec:applications}. In addition to our
main application of school choice in Chile, we study markets
in which feasibility constraints or objectives have a \emph{matroid}
structure.\footnote{Matroids are combinatorial objects with a particular structure. We
provide an overview of matroids in Section \ref{sec:math}.}
All omitted proofs are in the Appendix.

\section{Model}\label{sec:model}

Let $\mathcal{A}$ be a set of agents, $\mathcal{I}$ a set of institutions, and
$\mathcal{X}$ a finite set of bilateral contracts that specify a relation between
an agent and an institution. For each contract $x\in \mathcal{X}$, the
agent associated with the contract is denoted by $\alpha(x) \in \mathcal{A}$.
For each set of contracts $X \subseteq \mathcal{X}$
and agent $a \in \mathcal{A}$, let $X_a$ denote the set of contracts in $X$
that include $a$, i.e., $X_a=\{x\in X|\alpha(x)=a\}$. Likewise, for each
set of contracts $X \subseteq \mathcal{X}$ and institution $i\in \mathcal{I}$,
let $X_i$ denote the set of contracts in $X$ that include $i$.

Each agent $a\in \mathcal{A}$ has a preference relation (a linear order)
$R_a$ over $\mathcal{X}_a \cup\{\emptyset\}$, where $\emptyset$ is the null
contract representing the outside option or not being allocated any contract
in $\mathcal{X}$. The strict preference corresponding to $R_a$ is denoted
by $P_a$ and defined as, for $x,y \in \mathcal{X}_a \cup\{\emptyset\}$,
$x \mathrel{P}_a y$ if $x \mathrel{R_a} y$ and $x\neq y$. For agent
$a\in \mathcal{A}$, a contract $x\in \mathcal{X}_a \cup \{\emptyset\}$
is \emph{acceptable} if $x \mathrel{R_a} \emptyset$.
We write $R=(R_a)_{a\in \mathcal{A}}$ to denote a profile of preference relations for the agents. We also write $R_{-a}$ to denote $(R_{a'})_{a'\in \mathcal{A}\setminus\{a\}}$ for agent $a\in\mathcal{A}$.


A set of contracts $X_i\subseteq \mathcal{X}_i$ is called a
\emph{choice problem} for institution $i\in \mathcal{I}$. Given a choice problem,
the institution needs to choose a subset of the given set of contracts.
Therefore, each institution $i \in \mathcal{I}$ has a \emph{choice rule}.
More formally,
a choice rule $C_i: 2^{\mathcal{X}_i} \rightarrow 2^{\mathcal{X}_i}$ is
such that, for each $X_i\subseteq \mathcal{X}_i$, $C_i(X_i)\subseteq X_i$.
Let $\mathcal{C}_i$ denote the set of all choice rules for institution $i$.

A \emph{choice axiom} on $\mathcal{C}_i$ is a requirement on choice rules
for institution $i\in \mathcal{I}$, which may or may not be satisfied by
a given choice rule.
We introduce two choice axioms that play a
fundamental role in economic theory in general and, market design
and choice theory in particular. These choice axioms are central
to our results.

\begin{definition}
A choice rule $C_i \in \mathcal{C}_i$ satisfies \emph{path independence} if, for each $X_i,X_i'\subseteq \mathcal{X}_i$,
\[C_i(X_i \cup X_i')=C_i(C_i(X_i) \cup X_i').\]
\end{definition}

\cite{plott} introduces path independence as a rationality axiom
in a social choice context. In addition, it guarantees the existence
of a desirable outcome in two-sided markets.\footnote{Path independence is
equivalent to the conjunction of the \emph{substitutes} condition and the
\emph{irrelevance of rejected contracts} condition \citep{AizMal:1981AutCont}.}

\begin{definition}
A choice rule $C_i\in \mathcal{C}_i$ satisfies \emph{size monotonicity} if,
for each $X_i'\subseteq \mathcal{X}_i$ and $X_i\subseteq X_i'$,
\[|C_i(X_i)|\leq |C_i(X'_i)|.\]
\end{definition}

\cite{alkan03} introduce size monotonicity. It is also
known as the \emph{law of aggregate demand} in the matching with contracts
literature \citep{hatfi04}. It states that when the set of available contracts expands, at least as many contracts as before are chosen.

We fix the choice rule profile $C=(C_i)_{i\in \mathcal{I}}$
as well as the set of agents, institutions, and contracts throughout the paper.
Therefore, a \emph{matching problem} is simply a profile of preference
relations for the agents, $R=(R_a)_{a\in \mathcal{A}}$. Given a matching
problem, the goal is to find a \emph{matching}, which is defined as a
set of contracts $X\subseteq \mathcal{X}$ that contains at most one
contract for each agent, i.e., $|X_a|\leq 1$ for each $a\in \mathcal{A}$. Even
though agents can have at most one contract in a matching,
institutions can have contracts with several agents simultaneously.
Let $\mathcal{M}$ denote the set of all matchings.

A \emph{matching axiom} on $\mathcal{M}$ at a given problem $R$ is a requirement on matchings that may or may not be satisfied by a given matching.
A fundamental axiom in the mechanism design literature is that no
agent can be forced to sign an unacceptable contract.

\begin{definition}
A matching $X \subseteq \mathcal{X}$ satisfies \emph{individual rationality} if,
for each $a\in \mathcal{A}$,
\[X_a \mathrel{R_a} \emptyset.\]
\end{definition}

A \emph{matching rule} $\varphi$ associates each matching problem
$R=(R_a)_{a\in \mathcal{A}}$ with a matching $\varphi(R)\in \mathcal{M}$.
A matching rule $\varphi$ satisfies a matching axiom if, and only if, for
each matching problem $R$, $\varphi(R)$ satisfies the axiom.

The following axiom is a well-known axiom on matching rules, rather
than matchings, and requires that no agent can benefit from misreporting
their preferences.

\begin{definition}
A matching rule $\varphi$ is \emph{strategy-proof}, if, for each problem $R$,
agent $a\in \A$, and an alternative preference relation $R'_a$ for $a$,
\[\varphi_a(R) \mathrel{R_a} \varphi_a(R_a',R_{-a}).\]
\end{definition}

\section{Punctual Choice Axioms and Their Extensions}\label{sec:punext}
In this section, we identify a property of choice axioms that is central
to our analysis and introduce a method to extend such choice axioms to
matching axioms.

\subsection{Punctual Choice Axioms}\label{sec:punctual}
Consider an institution $i\in \mathcal{I}$. A choice axiom on $\mathcal{C}_i$ is
\emph{punctual} if it imposes a requirement on choice problems that can be checked
without referring to other choice problems. More
formally, a \emph{punctual choice axiom} is a correspondence
$\phi_i: 2^{\mathcal{X}_i} \rightrightarrows 2^{\mathcal{X}_i}$ where $\phi_i(X_i) \subseteq 2^{X_i}$ for each  $X_i\subseteq \mathcal{X}_i$, such that a choice rule $C_i \in \mathcal{C}_i$ \emph{satisfies} $\phi_i$ if, and only if, $C_i$ is a selection from $\phi_i$, i.e., for each $X_i\subseteq \mathcal{X}_i$,
\[C_i(X_i)\in \phi_i(X_i).\]

Punctual choice axioms have the following property, which is useful in verifying whether a given choice axiom is punctual: Let $i\in \mathcal{I}$ and $\phi_i$ be a punctual choice axiom on $\mathcal{C}_i$. Suppose that a choice rule $C_i$ for an institution $i\in \mathcal{I}$ is obtained by \emph{combining} two choice rules $C_i'$ and $C_i''$, i.e., for each $X_i\subseteq \mathcal{X}_i$, $C_i(X_i)=C_i'(X_i)$ or $C_i(X_i)=C''_i(X_i)$. Suppose also that both $C_i'$ and $C_i''$ satisfy $\phi_i$. Then, $C_i$ satisfies $\phi_i$ as well. In fact, punctual choice axioms are characterized by this property.

\begin{proposition}\label{prop:punctualchoice}
Let $i\in \mathcal{I}$. An axiom $\phi_i$ on $\mathcal{C}_i$ is punctual if and only if for any choice rule $C_i$ that is obtained by combining two choice rules $C_i'$ and $C_i''$ that satisfy $\phi_i$, $C_i$ also satisfies $\phi_i$.
\end{proposition}

To illustrate punctuality, we first explain that the two choice axioms introduced
in Section \ref{sec:model}, path independence and size monotonicity, are
not punctual. Then we consider examples of punctual choice axioms.

We start with path independence. Let $i\in \mathcal{I}$ be an institution
such that the set $\mathcal{X}_i$ includes two distinct contracts $x$ and $y$.
Consider the choice rule $C_i$ such that $C_i(X_i)=\emptyset$ for any problem
$X_i \subseteq \mathcal{X}_i$. Choice rule $C_i$ satisfies path independence and
$C_i(\{x\})=\emptyset$. Next consider choice rule $C'_i$ such
that $C'_i(X)=\{x\}$ for any problem $X_i \subseteq \mathcal{X}_i$ that
includes $x$ and, otherwise, $C'_i(X)=\emptyset$.
It is easy to check that choice rule $C'_i$ also satisfies path independence
and $C'_i(\{x,y\})=\{x\}$.
If path independence was punctual, then by Proposition \ref{prop:punctualchoice}, there
would exist a path-independent choice rule $C_i''$ such that $C_i''(\{x\})=C_i(\{x\})=\emptyset$ and $C_i''(\{x,y\})=C_i'(\{x,y\})=\{x\}$. However, such a choice rule does not
satisfy path independence because
\[C_i''\left(\{x,y\}\right)=\{x\}\neq C_i''\left(\{y\} \cup C_i''(\{x\})\right)=C_i''\left(\{y\}\right)\]
since $C_i''(\{y\})\subseteq \{y\}$ and $\{x\} \not \subseteq \{y\}$.
We conclude that path independence is not punctual.

To see that size monotonicity is not punctual, let $i\in \mathcal{I}$ be an institution
such that its set of contracts $\mathcal{X}_i$ include two distinct contracts $x$
and $y$.
Consider the choice rule $C_i$ such that $C_i(X_i)=\emptyset$ for any problem
$X_i \subseteq \mathcal{X}_i$. Choice rule $C_i$ satisfies size monotonicity and
$C_i(\{x,y\})=\emptyset$. Next consider choice rule $C'_i$ such
that $C'_i(X)=X$ for any problem $X_i \subseteq \mathcal{X}_i$.
Choice rule $C'_i$ also satisfies size monotonicity and $C'_i(\{x\})=\{x\}$.
If size monotonicity was punctual, then by Proposition \ref{prop:punctualchoice}, there
would exist a size-monotonic choice rule $C_i''$ such that $C_i''(\{x,y\})=C_i(\{x,y\})=\emptyset$ and $C_i''(\{x\})=C_i'(\{x\})=\{x\}$. However, such a choice rule does not satisfy size monotonicity because
\[\abs{C_i''(\{x,y\})}=0<1=\abs{C_i''(\{x\})}.\]
Therefore, size monotonicity is not punctual.

We now present two punctual choice axioms. These two axioms have been central in markets
where any agent and institution can be matched in a unique way, and, therefore, each agent-institution pair uniquely defines a contract, e.g., when matching students to public schools. In addition, these two axioms pertain to institutions whose choice behaviour is governed by a given \emph{capacity} and a \emph{priority ordering} over contracts.
Therefore, for each institution
$i\in \mathcal{I}$, let $\mathcal{X}_i$ be the set of all pairs including an agent and the institution, $q_i \in \mathbb{N}$ a capacity, and $\succ_i$ a priority
ordering (a linear order) over $\mathcal{X}_i$.

The first axiom, \emph{non-wastefulness},\footnote{Non-wastefulness is also called \emph{capacity filling} \citep{alkan03} and \emph{acceptance} \citep{kojman10}.} requires that, at each problem, the choice
rule never selects more contracts than the capacity and, furthermore,
does not reject any contract unless the capacity is reached.

\begin{definition}
A choice rule $C_i$ is \emph{non-wasteful} if, for each $X_i \subseteq \mathcal{X}_i$, \[\abs{C_i(X_i)}=\min\{|X_i|,q_i\}.\]
\end{definition}
Non-wastefulness is punctual because we can construct a correspondence so that a
choice rule satisfies this axiom if
and only if it is a selection from this correspondence. The
correspondence for non-wastefulness, denoted by $\phi_i^{nw}$, is given by, for each
$X_i \subseteq \mathcal{X}_i$,
\[
\phi_i^{nw}(X_i)=\{Y_i\subseteq X_i : |Y_i|=\min\{|X_i|,q_i\}\}.
\]



The second axiom, \emph{no justified envy},\footnote{No justified envy is also called priority compatibility \citep{echyen12}.} requires that no agent is chosen at the expense of a higher priority agent.

\begin{definition}
A choice rule $C_i$ satisfies \emph{no justified envy} if,
for each $X_i \subseteq \mathcal{X}_i$ and $x,y\in X_i$,
\[x\in C_i(X_i) \; \mbox{and} \; y\notin C_i(X_i) \; \implies \; x\succ_i y.\]
\end{definition}

No justified envy is a punctual choice axiom because it can be rewritten as
being a selection from the following correspondence, denoted by $\phi_i^{ne}$. For each
$X_i \subseteq \mathcal{X}_i$,
\[
\phi_i^{ne}(X_i)=\{Y_i\subseteq X_i : x\in Y_i \; \mbox{and} \; y\in X_i\setminus Y_i \; \implies \; x\succ_i y\}.
\]


\subsection{Extending Punctual Choice Axioms to Matchings}
In this section, we introduce a systematic way to extend any punctual
choice axiom to a matching axiom.

Given an institution $i\in \mathcal{I}$, a matching problem
$R=(R_a)_{a\in \mathcal{A}}$, and a matching $X\subseteq \mathcal{X}$,
define the \emph{demand} for $i$ at $X$ as the set of $i$'s
contracts that are preferred by
the agent associated with the contract to their assignment, more formally,
\[D_i(X)=\left\{x \in \mathcal{X}_i : x \mathrel{R_{\alpha(x)}} X_{\alpha(x)}\right\}.\footnote{With
a slight abuse of notation, for each agent $a\in \mathcal{A}$, we use
the preference relation $R_{a}$
to compare contracts of $a$ and sets that include
only one contract of $a$.}\]
By definition, $D_i(X)$ includes all contracts in $X_i$ because, for
each contract $x \in X_i$, we have $X_{\alpha(x)}=\{x\}$ since
$X$ is a matching and, hence, $x \mathrel{R_{\alpha(x)}} X_{\alpha(x)}$,
which implies that $x\in D_i(X)$.

Let $\phi_i$ be a punctual choice axiom on $\mathcal{C}_i$.
We define the \emph{extension} of $\phi_i$ to matchings, denoted
by $\overline{\phi}_i$, as follows: A matching $X \subseteq \mathcal{X}$
satisfies the matching axiom $\overline{\phi}_i$ at $R$
if,
\[X_i \in \phi_i(D_i(X)).\]
We can conceive the demand for institution $i$ at $X$, $D_i(X)$, as a
choice problem faced by $i$ and the set of contracts of $i$, $X_i$,
as the choice of $i$.
Therefore, $X$ satisfies the extended axiom at $R$ if, and only if,
it satisfies the choice axiom at choice problem $D_i(X)$.

Now we study the extensions of the two punctual choice axioms, non-wastefulness and
no justified envy, presented in
Section \ref{sec:punctual}. As before, for each institution $i\in \mathcal{I}$, let $\mathcal{X}_i$ be the set of all possible pairs of an agent and the institution,
$q_i \in \mathbb{N}$ a capacity, and $\succ_i$ a priority ordering (a linear order)
over $\mathcal{X}_i$.

The extension of non-wastefulness can be
stated as, for each $X_i \subseteq \mathcal{X}_i$,
\[X_i\in \phi_i^{nw}(D_i(X))=\{Y_i\subseteq D_i(X) : |Y_i|=\min\{|D_i(X)|,q_i\}\}.\]
Since $X_i\subseteq D_i(X)$, the extension of non-wastefulness is equivalent to
\[|X_i|=\min\{|D_i(X)|,q_i\}.\]
First, non-wastefulness implies that $|X_i|\leq q_i$. Furthermore, it is
trivially satisfied if $|X_i|=q_i$ since $D_i(X)\supseteq X_i$
and, hence, $|D_i(X)|\geq |X_i|=q_i$.
Otherwise, if $|X_i|<q_i$, it is equivalent to $X_i=D_i(X)$. Therefore,
if $|X_i|<q_i$, there cannot exist
a contract $x\in \mathcal{X}_i$ such that $x \mathrel{P_{\alpha(x)}} X_{\alpha(x)}$.
Thus, the extension of non-wastefulness to matchings can be written as follows.
\begin{definition}
A matching $X\subseteq \mathcal{X}$ satisfies
\emph{non-wastefulness for institution $i\in \mathcal{I}$} if $|X_i|\leq q_i$, and
\[x\in \mathcal{X}_i \mbox{ and } x \mathrel{P_{\alpha(x)}} X_{\alpha(x)} \implies |X_i|=q_i.\]

We say $X$ satisfies \emph{non-wastefulness} if it is non-wasteful for
every institution.
\end{definition}

Next we consider no justified envy.\footnote{No justified envy was introduced by \cite{abdulson03} in the context of school choice.} The extension of this axiom
to matchings is, for each $X_i\subseteq \mathcal{X}_i$,
\[X_i\in \phi_i^{ne}(D_i(X))=\{Y_i\subseteq D_i(X) : x\in Y_i \; \mbox{and} \; y\in D_i(X)\setminus Y_i  \implies  x \succ_i y\}.\]
Since $X_i \subseteq D_i(X)$, we can rewrite this as
\[x\in X_i \; \mbox{and} \; y\in D_i(X)\setminus X_i \; \implies \; x \succ_i y.\]
Note that $y\in D_i(X)\setminus X_i$ is equivalent to
$y \mathrel{P_{\alpha(y)}} X_{\alpha(y)}$.
Therefore, every agent who prefers their contract with $i$ to their assigned
contract in $X$ should have a lower priority than any agent who has a contract
with $i$.

\begin{definition}
A matching $X\subseteq \mathcal{X}$ satisfies \emph{no justified envy
at institution $i$} if, there exist no $x\in \mathcal{X}_i \setminus X$ and
$y\in X$ such that $x \mathrel{P_{\alpha(x)}} X_{\alpha(x)}$ and $x \succ_i y$.
We say $X$ satisfies \emph{no justified envy} if it satisfies no justified envy at every institution.
\end{definition}

\section{Deferred-Acceptance Rule and Its Characterization}\label{sec:DA}
Let $C=(C_i)_{i\in \mathcal{I}}$ be a choice rule profile for institutions. The
deferred-acceptance rule based on $C$ is defined through the following algorithm.

\begin{quote}
	\begin{center}
		{\bf Deferred-Acceptance Algorithm Based on $C$}
	\end{center}
	
	\noindent {\bf  Step 1:} Each agent proposes their most preferred acceptable contract to the associated institution (if there is none, they are assigned the null contract). Each institution $i$ considers proposals (from this step), say $X^1_i$, and tentatively accepts $C_i(X^1_i)$ and permanently rejects
$X^1_i \setminus C_i(X^1_i)$. If there is no rejection by any institution at this step, then stop and return the resulting matching. Otherwise, go to Step 2.
	
	\noindent {\bf Step $s\geq 2$:} Each agent whose contract was rejected
in the previous step proposes their next most preferred acceptable contract to the associated institution (if there is none, they are assigned the null contract). Each institution $i$ considers its tentatively accepted contracts from the previous step together with proposals from this step, say $X^s_i$, and tentatively accepts  $C_i(X^s_i)$ and permanently rejects $X^s_i \setminus C_i(X^s_i)$. If there is
no rejection by any institution at this step, then stop and return the
resulting matching.
	Otherwise, go to Step $s+1$.
	\end{quote}
The algorithm must stop at a finite step because no agent proposes any
contract more than once and the number of contracts is finite. The DA rule based on $C$ chooses, at each problem, the matching defined by the acceptances at the last step of the DA algorithm based on $C$.

We are now ready to present our main result, which extends a characterization of institutional choice rules to a characterization of the DA rule based on these choice rules.

\begin{theorem}\label{thm:main}
Suppose that, for each institution $i\in I$, the choice rule $C_i$ is characterized by a set of punctual axioms $\Phi_i$ and, furthermore, satisfies path independence and size monotonicity.\footnote{In other words, $C_i$ is the unique
choice rule in $\mathcal{C}_i$ that satisfies all axioms in $\Phi_i$ simultaneously and, furthermore, these axioms together imply path independence and size monotonicity.}
Then, a matching rule satisfies
individual rationality, strategy-proofness, and the extensions of
$(\Phi_i)_{i\in \mathcal{I}}$ if, and only if, it is the $DA$ rule based on $C$.
\end{theorem}

Therefore, any characterization of path-independent and size-monotonic choice rules by punctual choice axioms yields a characterization of the
deferred-acceptance rule in which institutions use these choice rules.\footnote{A stronger statement would start with the supposition that for each institution $i\in I$, $C_i$ is characterized by path independence, size monotonicity, and a set of punctual axioms. In Appendix \ref{stronger_thm}, we show that the corresponding stronger
statement does not necessarily hold.}
This is especially important for market design because policymakers can be
presented with axioms as properties of matching rules so that they can decide
how to design centralized clearinghouses such as the medical residency match or school choice systems around the world.

Now we provide a simple illustration of Theorem \ref{thm:main} and derive
the deferred-acceptance rule characterization by \cite{albar94} as
a corollary of Theorem \ref{thm:main}. We present the main applications
in Section \ref{sec:applications}.

As in Section \ref{sec:punctual}, for each institution $i\in \mathcal{I}$,
let $\mathcal{X}_i$ be the set of all pairs of an agent and the institution,
$q_i \in \mathbb{N}$ a capacity, and $\succ_i$ a priority ordering
(a linear order) over $\mathcal{X}_i$. For concreteness, suppose that
agents are students and institutions are schools.

For each school $i$, let $C^r_i$ be the following \emph{responsive} choice rule: at
each choice problem, contracts with the highest priority at $i$ are chosen until
the capacity $q_i$ is reached or no contract is left. It is known that a choice
rule satisfies no justified envy and non-wastefulness if, and only if, it is
the responsive choice rule $C^r_i$, and, furthermore, $C^r_i$ satisfies
path independence and size monotonicity \citep{chayen18}.\footnote{More
precisely, \cite{chayen18} characterize the responsive choice rules
using the \emph{weaker axiom of revealed preference}, which is a
non-punctual axiom. However, in our setting, it is easy to see that
no justified envy can replace the weaker axiom of revealed preference.}

The following characterization of the DA rule based on responsive choice rules
follows as a corollary to our Theorem \ref{thm:main}.\footnote{To be more
precise, \cite{albar94} show that $DA$ rule based on $(C_i^r)_{i\in \mathcal{I}}$
is characterized by
\emph{stability} and strategy-proofness. Moreover, stability is equivalent
to individual rationality, non-wastefulness, and no justified envy \citep{balson99}.}

\begin{corollary}[\citeauthor{albar94},\citeyear{albar94}]\label{coro:albar94}
The $DA$ rule based on $(C_i^r)_{i\in \mathcal{I}}$ is characterized by
non-wastefulness, no justified envy, individual rationality, and strategy-proofness.
\end{corollary}

\section{Applications}\label{sec:applications}
Applications of our theory include school choice in Chile as well as matching
problems in which feasibility constraints at institutions have a matroid structure
and matching problems in which institutions have matroidal objectives. We
first provide mathematical definitions on matroids \citep{oxley} that we need
for completeness and then we present our applications.

\subsection{Mathematical Preliminaries}\label{sec:math}

A \emph{matroid} is a pair $(\mathcal{X},\mathcal{F})$ where $\mathcal{X}$ is a finite set and $\mathcal{F}$ is a collection of subsets of $\mathcal{X}$ such that
\begin{enumerate}
\item[(I1)] $\emptyset \in \mathcal{F}$,

\item[(I2)] If $X'\in \mathcal{F}$ and $X\subseteq X'$, then $X\in \mathcal{F}$, and

\item[(I3)] If $X, X'\in \mathcal{F}$ and $|X|<|X'|$, then there exists
$x\in X'\setminus X$ such that $X\cup \{x\}\in \mathcal{F}$.
\end{enumerate}


Each set in $\mathcal{F}$ is called an \emph{independent set}. A set $X \subseteq \mathcal{X}$ is a \emph{base} of matroid $(\mathcal{X},\mathcal{F})$ if $X \in \mathcal{F}$ and there is no $Y \in \mathcal{F}$ such that $X\subsetneq Y$. By \emph{I3}, all bases of a matroid have the same cardinality. The \emph{rank} of $X\subseteq \mathcal{X}$, denoted by $r(X)$, is defined as the cardinality of
any of maximal independent subset of $X$, which is well-defined by \emph{I3}.
The rank function of a matroid has the following properties:
\begin{enumerate}
  \item[(R1)] If $X \subseteq \mathcal{X}$, then $0 \leq r(X) \leq |X|$.
  \item[(R2)] If $X \subseteq X' \subseteq \mathcal{X}$, then $r(X) \leq r(X')$.
  \item[(R3)] If $X,X' \subseteq \mathcal{X}$, then
            \[r(X\cup X') + r(X\cap X')\leq r(X)+r(X').\]
  \end{enumerate}

Given a matroid $(\mathcal{X},\mathcal{F})$ and a priority ordering $\succ$ over
$\mathcal{X}$, the \emph{greedy choice rule} on $\mathcal{X}$,
$C^g:2^{\mathcal{X}} \rightarrow 2^{\mathcal{X}}$, is defined as follows.

\begin{quote}
	\noindent{}{\bf Greedy Choice Rule} {\boldmath$C^{g}$}\\
	Let $X\subseteq \mathcal{X}$. Set $C^g_0(X)=\emptyset$.
	
	\noindent{}{\bf Steps $s\in \{1,\ldots ,|X|\}$:} Consider the $s$-th highest priority element in $X$, say $x_s$. If $C^g_{s-1}(X) \cup \{x_s\}\in \mathcal{F}$,
then set $C^g_s(X)=C^g_{s-1}(X)\cup \{x_s\}$. Otherwise, if $C^g_{s-1}(X) \cup \{x_s\}\notin \mathcal{F}$, then set $C^g_s(X)=C^g_{s-1}(X)$.
\end{quote}

For each $X \subseteq \mathcal{X}$, define $C^g(X)=C^g_{|X|}(X)$.

\subsection{Chilean School Choice: Enrollment Guarantees with
Overlapping Reserves}\label{sec:Chile}
In this application, we consider the school choice setting in Chile.
The \emph{School Inclusion Law} was promulgated in 2015. Since then a centralized
clearinghouse based on the deferred-acceptance algorithm has been implemented
nationwide for K-12 student assignment \citep{correa2022}.

At each school, according to the law, returning students have enrollment guarantees and other students are prioritized depending on some criteria such as whether they
have a sibling attending the school or a parent working at the school. In addition, there are minimum-guarantee reservations for students with disabilities, economically
disadvantaged students, and students with high academic performance.

To embed Chilean school choice in our setting, suppose that agents are
students and institutions are schools. A student can be assigned to a
school in only one possible way.\footnote{Students cannot be treated
differently under the School Inclusion Law.} Therefore, let
the set of contracts be the set of all possible student-school
pairs, i.e., $\mathcal{X}=\mathcal{A}\times \mathcal{I}$. For each
school $i \in \mathcal{I}$, there exist a priority relation $\succ_i$ on
$\mathcal{X}_i$, a capacity $q_i \in \mathbb{N}$, and a set of
returning students with cardinality of at most $q_i$. If a returning
student applies to the school, then the school has to admit the student.
For each school $i \in \mathcal{I}$, we denote the set of contracts of
$i$ with its returning students by $\mathcal{X}_i^{ge} \subseteq \mathcal{X}_i$.

There exist a set of types $\mathcal{T}$, a type function
$\tau: \mathcal{A} \rightarrow 2^{\mathcal{T}}$, and a reserve profile
$(r_i^t)_{t\in \mathcal{T}}$ for each school $i\in \mathcal{I}$. Therefore,
for each student $a\in \mathcal{A}$, $\tau(a)\subseteq \mathcal{T}$ is the
set of traits that the student has.
For each school $i\in \mathcal{I}$ and type $t\in \mathcal{T}$, $r_i^t$
represents the number of seats reserved for type-$t$ students at school $i$.
Therefore, from any set of applicants, $i$ has to admit at least $r_i^t$
students of type $t$ if there are more than $r_i^t$ type-$t$ applicants and,
otherwise, $i$ has to admit all type-$t$ students.
In Chile, there are reservations for three groups: students with
disabilities, economically disadvantaged students, and students
with high academic performance.

Consider an economically disadvantaged student who has a high academic performance. When the student is admitted to a school, the student can be counted as an economically disadvantaged student or a student with high academic performance or both.
We assume that each student accounts for only one of the traits that they
have.\footnote{The same assumption is made implicitly in \cite{correa2022} since
each student is assigned a seat in their choice rule construction. See also
\cite{sonmez/yenmez:22}.}
Given a set of students that a school admits, we need to determine how many
reserved positions are filled. To this end, we construct a \emph{reserve graph} with
students on the one side and reserved seats for the school on the other side. A
student is connected with a seat reserved for a trait if the student has the
trait.

For each school $i\in \mathcal{I}$, construct $\mathcal{F}_i$ as follows:
$X_i \subseteq \mathcal{X}_i$ is in $\mathcal{F}_i$ if the set of students
who have contracts in $X_i$ can be assigned to different positions in the
reserve graph so that each student is assigned to a node that they are
connected to. $(\mathcal{X}_i,\mathcal{F}_i)$ is a well-known matroid called
the \emph{transversal matroid}. Let the rank function of this matroid be
denoted as $r_i$.\footnote{See Section \ref{sec:math} for the
rank function properties.} Then, for any $X_i\subseteq \mathcal{X}_i$,
$r_i(X_i)$ is the maximum number of reserved positions that can be assigned to
students who have contracts in $X_i$.

Now we are ready to identify choice rule properties in spirit of the School
Inclusion Law in Chile. The first property is non-wastefulness that we define
above. For the rest of the properties, fix a school $i \in \mathcal{I}$.

\begin{definition}
Choice rule $C_i$ satisfies \emph{guaranteed enrollment for returning students} if,
for each $X \subseteq \mathcal{X}$,
\[
\mathcal{X}_i^{ge} \cap X_i \subseteq C_i(X).
\]
\end{definition}
This property requires schools to accept returning students.

\begin{definition}
Choice rule $C_i$ satisfies \emph{maximal utilization of reservations} if,
for each $X \subseteq \mathcal{X}$ and $x\in X_i \setminus C_i(X_i)$,
\begin{enumerate}
\item [(i)] if $\abs{C_i(X_i)}<q_i$, then $r_i(X_i\cup \{x\})=r_i(X_i)$, and
\item [(ii)] if $\abs{C_i(X_i)}=q_i$, then for any
$y \in C_i(X_i)\setminus \mathcal{X}_i^{ge}$,
    \[r_i\left((C_i(X_i)\setminus\{y\})\cup\{x\}\right) \leq r_i(C_i(X_i)).\]
\end{enumerate}
\end{definition}
Since $r_i$ gives the number of reserved seats utilized, this condition
states that a rejected student cannot increase the reserve utilization
of the school either by taking an empty seat or by replacing another
student who does not have enrollment guarantees.

The following is a generalization of the no justified envy condition we define
in Section \ref{sec:punctual} and uses the reserve utilization to determine when the
envy of a student is justified.
\begin{definition}
Choice rule $C_i$ satisfies \emph{no justified envy under reserves} if,
for each $X\subseteq \mathcal{X}$, $x\in C_i(X)\setminus \mathcal{X}_i^{ge}$,
and $y\in X_i \setminus C_i(X_i)$,
\[
y \succ_i x \implies r_i((C_i(X_i)\setminus\{x\})\cup\{y\})<r_i(C_i(X_i)).
\]
\end{definition}

In plain words, if a student who does not have enrollment guarantees is accepted at the expense of a higher priority student, then it must be that accepting the higher priority student instead decreases the reserve utilization at the school.


\medskip
We are now ready to define our choice rule.
\begin{quote}
        \noindent{}{\bf Guaranteed-Enrollment Choice Rule} {\boldmath$C^{ge}_{i}$}\\
        \noindent{}{\bf Step 1:} Let $X_i \subseteq \mathcal{X}_i$ be a set of
        contracts associated with school $i\in \mathcal{I}$ and
        $X^0_i = X_i \cap \mathcal{X}_i^{ge}$.
        If $|X^0_i|=q_i$, then return $X^0_i$ and stop the procedure.
	
		\noindent{}{\bf Step 2.s:} {\boldmath($s\in \{1, \ldots, \sum_{t\in \mathcal{T}} r^t_i\}$)}:
		Assuming such a contract exists,
		choose the highest priority contract $x^s \in X_i \setminus X^{s-1}_i$
		such that
		 \[r_i(X^{s-1}_i \cup \{x^s\})=r_i(X^{s-1}_i)+1\]
		and let $X_i^s=X_i^{s-1}\cup \{x^s\}$.
		If $|X_i^s|=q_i$, then return $X_i^s$ and stop the procedure.
		If no such contract exists, proceed to Step 3.

        \noindent{}{\bf Step 3:}
           For the remaining unfilled seats, choose unassigned contracts with
           the highest priority either until all seats are filled or all contracts are selected. Return all chosen contracts.
\end{quote}
The guaranteed-enrollment choice rule has three steps. At the first step,
contracts of returning
students are chosen. At the second step, contracts of students who increase the
reserve utilization are added. Finally, at the third step, remaining contracts are
chosen to fill empty seats.

\begin{proposition}\label{proposition_GE}
Consider a school $i\in \mathcal{I}$. A choice rule $C_i$ satisfies
\begin{itemize}
\item[(i)] guaranteed enrollment for returning students,
\item[(ii)] maximal utilization of reservations,
\item[(iii)] no justified envy under reserves, and
\item[(iv)] non-wastefulness
\end{itemize}
if, and only if, it is the guaranteed-enrollment choice rule $C^{ge}_{i}$.
Furthermore, $C^{ge}_{i}$ satisfies path independence and size monotonicity.
\end{proposition}

Next we introduce matching axioms in spirit of the law. In the Appendix, we show
that the choice axioms above are punctual and their extensions are the matching
axioms that we define below.

\begin{definition}
A matching $X\subseteq \mathcal{X}$ satisfies \emph{guaranteed enrollment for
returning students} if, for each $i \in \mathcal{I}$ and
$x\in \mathcal{X}^{ge}_i \setminus X_i$ we have
\[X_{\alpha(x)} \mathrel{P_{\alpha(x)}} x.\]
\end{definition}
Guaranteed enrollment for returning students requires each returning student
to be either matched with the school they are returning to or matched with a school that
they prefer more.

\begin{definition}
A matching $X\subseteq \mathcal{X}$ satisfies
\emph{maximal utilization of reservations} if,
for each $i\in \mathcal{I}$ and $x \in \mathcal{X}_i \setminus X_i$ with
$x \mathrel{P_{\alpha(x)}} X_{\alpha(x)}$,
\begin{enumerate}
\item [(i)] if $\abs{X_i}<q_i$, then $r_i(X_i\cup \{x\})=r_i(X_i)$, and
\item [(ii)] if $\abs{X_i}=q_i$, then for any
$y \in X_i \setminus \mathcal{X}_i^{ge}$,
    \[r_i\left((X_i \setminus\{y\})\cup\{x\}\right) \leq r_i(X_i).\]
\end{enumerate}
\end{definition}
In plain words, for any school $i$, a student who strictly prefers school $i\in \mathcal{I}$
to their assignment cannot take an empty seat or replace a student assigned to
$i$, who is not a returning student, and increase the number of reserved positions
that can be filled at $i$.

\begin{definition}
A matching $X\subseteq \mathcal{X}$ satisfies \emph{no justified envy under reserves}
if, for each $i\in I$,
$x \in \mathcal{X}_i \setminus X_i$ with $x \mathrel{P_{\alpha(x)}} X_{\alpha(x)}$,
and $y \in X_i\setminus \mathcal{X}_i^{ge}$
\[
x \succ_i y \Longrightarrow r_i((X_i\setminus\{y\})\cup\{x\})<r_i(X_i).
\]
\end{definition}
Equivalently, if a student who does not gave enrollment guarantees is matched with a school
at the expense of a higher priority student, then it must be that admitting the higher priority student instead decreases the reserve utilization at the school.



By applying Theorem \ref{thm:main} to Proposition \ref{proposition_GE} we get the following
result.

\begin{corollary}\label{cor:ge}
A matching rule satisfies
\begin{itemize}
  \item[(i)] guaranteed enrollment for returning students,
  \item[(ii)] maximal utilization of reservations,
  \item[(iii)] no justified envy under reserves,
  \item[(iv)] non-wastefulness,
  \item[(v)] individual rationality, and
  \item[(vi)] strategy-proofness,
\end{itemize}
if, and only if, it is $DA$ based on $C^{ge}$.
\end{corollary}

\subsection{Matroid Constraints and Greedy Rule}
In this section, we consider applications in which institutions
have feasibility constraints that form a matroid.

Consider an institution $i\in \mathcal{I}$. Let $\mathcal{F}_i$ denote
the collection of set of contracts that satisfy the feasibility
constraints at $i$. In other words, a subset of $\mathcal{X}_i$ satisfies
the feasibility constraints at $i$ if, and only if, it is a member of
$\mathcal{F}_i$.  We assume that $(\mathcal{X}_i,\mathcal{F}_i)$ is a matroid.\footnote{See Section \ref{sec:math} for the definition of a matroid.}

An example of feasibility constraints that form a matroid is when
institutions have type-specific quotas. More formally, let $\mathcal{T}$ be
a set of agent types and $\tau: \mathcal{A} \rightarrow \mathcal{T}$ a
type function so that the type of agent $a \in \mathcal{A}$ is
$\tau(a)\in \mathcal{T}$.
Each institution $i \in \mathcal{I}$ has a capacity $q_i\in \mathbb{N}$ and,
for each type $t\in \mathcal{T}$, a quota $q_i^t \in \mathbb{N}$ for
type-$t$ agents. Therefore, in a feasible set, the number of type-$t$
agents at institution $i$ cannot be more than $q_i^t$.
Suppose that each agent can be matched with an institution
in a unique way, so $\mathcal{X}=\mathcal{I}\times \mathcal{A}$. Hence,
the collection of feasible sets at an institution $i\in \mathcal{I}$ can
be written as
\[\mathcal{F}_i=\{X_i \subseteq \mathcal{X}_i : |X_i|\leq q_i \text{ and, for
each } t\in \mathcal{T}, \abs{\{x\in X_i:\tau(\alpha(x))=t\}} \leq q_i^t\}.\]
This example can be generalized so that agents can have multiple types and
$(\mathcal{X}_i,\mathcal{F}_i)$ is a transversal matroid.

Now we consider desirable properties of choice rules in this context.

\begin{definition}
Choice rule $C_i$ satisfies \emph{feasibility} if, for each $X \subseteq \mathcal{X}$,
\[
C_i(X) \in \mathcal{F}_i.
\]
\end{definition}
The chosen set must be feasible.

\begin{definition}
Choice rule $C_i$ satisfies \emph{rank maximality} if, for each
$X \subseteq \mathcal{X}$,
\[
r_i(C_i(X))= r_i(X).
\]
\end{definition}
Rank maximality requires that as long as the feasibility constraint is not violated,
the institution chooses as many contracts as possible. More specifically, if a contract
$x$ is rejected, adding it to the chosen set must violate feasibility (i.e.,
$C_i(X) \cup\{x\} \notin\mathcal{F}_i$).



\begin{definition}
Choice rule $C_i$ satisfies \emph{no justified envy under rank} if,
for each $X \subseteq \mathcal{X}$, $x \in C_i(X)$, and $y \in X_i \setminus C_i(X)$,
\[
y\succ_i x \Longrightarrow r_i\left((C_i(X)\setminus\{x\})\cup\{y\}\right)<r_i(C_i(X)).
\]
\end{definition}
This is a different version of the axiom
\emph{no justified envy under reserves} that we study in Section \ref{sec:Chile}
because there are no returning students who have guaranteed enrollment
in the current setting. Furthermore, the rank function in the previous section
corresponds to a transversal matroid so that for any set the rank is equal to the
number of reserved positions that can be filled whereas in this section the rank
of a set can have other interpretations.

The following is a characterization of the greedy rule introduced in
Section \ref{sec:math}.

\begin{proposition}[\citeauthor{sonmez/yenmez:22}, \citeyear{sonmez/yenmez:22}]
A choice rule $C_i$ satisfies
\begin{itemize}
  \item[(i)] feasibility,
  \item[(ii)] rank maximality, and
  \item[(iii)] no justified envy under rank
\end{itemize}
if, and only if, it is the greedy choice rule $C_i^g$.
\end{proposition}

Next we consider the following matching axioms that are desirable in this setting.
In the appendix, we show that each matching axiom introduced below is the extension
of the matching axiom with the same name defined above.

\begin{definition}
A matching $X\subseteq \mathcal{X}$ satisfies \emph{feasibility} if, for
each $i\in \mathcal{I}$
\[X_i \in \mathcal{F}_i.\]
\end{definition}
For each institution the outcome must be feasible.

\begin{definition}
A matching $X\subseteq \mathcal{X}$ satisfies \emph{rank maximality} if,
for each $i\in \mathcal{I}$ and $x \in \mathcal{X}_i \setminus X_i$ with
$x \mathrel{P_{\alpha(x)}} X_{\alpha(x)}$,
\[
r_i\left(X_i\cup\{x\}\right)= r_i(X_i).
\]
\end{definition}
Rank maximality states that no contract of an agent who
prefers this contract to their current assignment can be added without
violating feasibility.

\begin{definition}
A matching $X\subseteq \mathcal{X}$ satisfies \emph{no justified envy under
rank} if,
for each $i\in I$, $x\in X_i$, and
$y\in \mathcal{X}_i \setminus X_i$ with $y \mathrel{P_{\alpha(y)}} X_{\alpha(y)}$,
\[
y\succ_i x \Longrightarrow r_i\left((X_i\setminus\{x\})\cup\{y\}\right)<r_i(X_i).
\]
\end{definition}

In plain words, if a student is matched with a school at the expense of a higher priority student,
then it must be that admitting the higher priority student instead lowers the rank at
the institution.

As a corollary of Theorem \ref{thm:main} we get the following.

\begin{corollary}\label{cor:mat1}
A matching rule satisfies
\begin{itemize}
  \item[(i)] feasibility,
  \item[(ii)] rank maximality,
  \item[(iii)] no justified envy under rank,
  \item[(iv)] individual rationality, and
  \item[(v)] strategy-proofness,
\end{itemize}
if, and only if, it is $DA$ based on $G$ where $G=(G_i)_{i\in \mathcal{I}}$ is the profile of greedy rules for institutions.
\end{corollary}

\subsection{Matroidal Objectives with Non-wastefulness}\label{sec:nw-matroid}
Suppose that each institution $i\in \mathcal{I}$ has a
capacity $q_i \in \mathbb{N}$ and objectives that can be represented
by the rank function $r_i$ of a matroid $(\mathcal{X}_i,\mathcal{F}_i)$,
where $\mathcal{F}_i$ is a collection of subsets of $\mathcal{X}_i$ such
that each set in $\mathcal{F}_i$ has cardinality of at most $q_i$.
Given a set of contracts $X_i \subseteq \mathcal{X}_i$, the extent to
which $X_i$ achieves the matroidal objectives is measured by the rank function
$r_i$ of matroid $(\mathcal{X}_i,\mathcal{F}_i)$. In economic applications,
these matroidal objectives typically correspond to some distributional
policies.\footnote{For example, see
\citet{kamada_kojima_aer, kamada_kojima_jet, kamada_kojima_te} who study
matching problems with regional constraints.}
Suppose also that each $i\in \mathcal{I}$ has a priority order (linear order)
$\succ_i$ over $\mathcal{X}_i$.

\begin{definition}
A choice rule $C_i$ satisfies \emph{matroidal objectives} if for each
$X \subseteq \mathcal{X}_i$ and each $X'\subseteq X$,
\[r_i(C_i(X)) \geq r_i(X').\]
\end{definition}
The chosen set of contracts must be an optimal subset of applications.

Now, we identify the choice rule characterized by the above axiom together with
non-wastefulness and no justified envy under rank. Let
$X_i \subseteq \mathcal{X}_i$ be a set of contracts considered at
institution $i\in \mathcal{I}$.
\begin{quote}
        \noindent{}{\bf Non-wasteful Matroid Choice Rule} {\boldmath$C^{m}_{i}$}\\
		\noindent{}{\bf Step 1:} Apply the greedy rule for matroid $(\mathcal{X}_i,\mathcal{F}_i)$ on $X_i$.

        \noindent{}{\bf Step 2:}
           For unfilled seats, choose unassigned contracts with the highest priority
           until either all seats are filled or all contracts are selected. Return
           the set of all chosen contracts.\footnote{This choice rule
           generalizes \emph{meritorious horizontal choice rule} in
           \cite{sonmez/yenmez:22} where each institution is endowed with a
           transversal matroid.}
\end{quote}

Next we consider the following property of a matching. In the Appendix, we show that
this property is the extension of the choice rule axiom with the same name defined above.

\begin{definition}
A matching $X\subseteq \mathcal{X}$ satisfies \emph{matroidal objectives} if,
for each $i\in I$ and $x\in \mathcal{X}_i \setminus X_i$ with
$x \mathrel{P_{\alpha(x)}} X_{\alpha(x)}$,
\[
r_i(X_i \cup \{x\}) = r_i(X_i).
\]
\end{definition}
A matching satisfies matroidal objectives if no institution can
increase the rank of its contracts with a student who prefers the
institution to their current match.

\begin{proposition}\label{proposition_matroid}
A choice rule $C_i$ satisfies
\begin{itemize}
\item[(i)] matroidal objectives,
\item[(ii)] no justified envy under rank, and
\item[(iii)] non-wastefulness
\end{itemize}
if, and only if, it is the matroid choice rule $C^m_{i}$. Furthermore,
$C^m_{i}$ satisfies path independence and size monotonicity.
\end{proposition}

We get the following corollary as an application of Theorem \ref{thm:main}.

\begin{corollary}\label{cor:mat2}
A matching rule satisfies
\begin{itemize}
\item[(i)] non-wastefulness,
\item[(ii)] no justified envy under rank,
\item[(iii)] individual rationality, and
\item[(iv)] strategy-proofness,
\end{itemize}
if, and only if, it is $DA$ based on $C^m_{i}$.
\end{corollary}

\section{Conclusion}

Our main result shows how to extend any characterization of path-independent and size-monotonic choice rules by punctual choice axioms to a characterization of the
DA rule in which priorities at institutions are determined by these choice rules.
By identifying a set of properties that characterizes the DA rule, we anticipate
that our method can be effective in persuading policymakers to establish
centralized clearinghouses similar to the medical residency match in the US
and school choice programs around the world.


Our work establishes an intimate relationship between choice rules and matching
rules that have been studied separately with a few notable exceptions and
identifies punctual choice axioms as the key property in this relationship.
Making such connections may help us advance different branches of economics by better
understanding the underlying theories.

Our results are based on two fundamental choice axioms that have
been central in the market design literature: path independence and size monotonicity.
Recent work has focused on weakening these properties.\footnote{For example,
see \cite{hatkoj10}, \cite{hatkom14}, and \cite{hatfield_et_al_2021}.}
We anticipate that our main insights may hold under weaker conditions, which we
leave for future research.

\bibliographystyle{chicago}
\bibliography{matching}
\newpage

\appendix

\section{Proof of Proposition \ref{prop:punctualchoice}}
\textit{Only if part:} Since $C'_i$ satisfies the punctual axiom $\phi_i$, for each
$X_i \subseteq \mathcal{X}_i$, $C'_i(X_i) \in \phi_i(X_i)$.
Likewise, since $C''_i$ satisfies the punctual axiom $\phi_i$,
$C''_i(X_i) \in \phi_i(X_i)$.

Therefore, any choice rule $C_i$ such that, for any
$X_i\subseteq \mathcal{X}_i$, $C_i(X_i)=C'_i(X_i)$ or $C_i(X_i)=C''_i(X_i)$
has the property that $C_i(X_i) \in \phi_i(X_i)$. We conclude that
any such choice rule $C_i$ satisfies $\phi_i$.

\textit{If part:} Let $\phi_i$ be an arbitrary choice axiom on $\mathcal{C}_i$. Suppose that any choice rule $C_i$ which is obtained by combining two choice rules $C_i'$ and $C_i''$ that satisfy $\phi_i$, also satisfies $\phi_i$. We construct a correspondence $\phi_i: 2^{\mathcal{X}_i} \rightrightarrows 2^{\mathcal{X}_i}$ as follows. For each $X_i\subseteq \mathcal{X}_i$, $X_i'\in \phi_i(X_i)$ if and only if there exists a choice rule $C_i$ that satisfies $\phi_i$ and $C_i(X_i)=X_i'$. We show that a choice rule satisfies the axiom $\phi_i$ if and only if it is a selection from the correspondence $\phi_i$, which implies
that $\phi_i$ is punctual. The fact that any choice rule $C_i$ that satisfies $\phi_i$ is a selection from the correspondence $\phi_i$ immediately follows from the construction of the correspondence $\phi_i$.

To show the other direction, first define the \emph{distance} between any two choice rules $C_i$ and $C_i'$ as the number of choice problems at which they differ, i.e., \[d(C_i,C_i')=|\{X_i\subseteq \mathcal{X}_i: C_i(X_i)\neq C_i'(X_i)\}|.\]
Now, let $C_i$ be an arbitrary choice rule that is a selection from the correspondence $\phi_i$. Suppose that $C_i$ does not satisfy the axiom $\phi_i$. Let $C'_i$ be a closest choice rule to $C_i$ that satisfies the axiom $\phi_i$, i.e., $C_i'$ satisfies $\phi_i$ and there is no other choice rule $C_i''$ satisfying $\phi_i$ such that $d(C_i,C_i'')<d(C_i,C_i')$ (since there are finitely many choice problems, such a choice rule exists). Let $X_i$ be any problem with $C_i(X_i)\neq C'_i(X_i)$. Since $C_i$ is a selection from $\phi_i$, there exists a choice rule $C_i''$ that satisfies the axiom $\phi_i$ and $C_i''(X_i)=C_i(X_i)$. Now, consider the choice rule $C_i'''$ defined as follows: $C_i'''(X_i)=C_i''(X_i)=C_i(X_i)$, and for any other problem $X_i'$, $C_i'''(X_i')=C_i'(X_i')$. Note that $C_i'''$ combines $C_i'$ and $C_i''$ that satisfy $\phi_i$, and by the only if part above, $C_i'''$ satisfies $\phi_i$ as well. Also note that $d(C_i,C_i''')=d(C_i,C_i')-1$, contradicting to the assumption that $C'_i$ is a closest choice rule to $C_i$ that satisfies the axiom $\phi_i$.

\bigskip

\section{Proof of Theorem \ref{thm:main}}
\textit{If part:} We first show that $DA$ based on $C$ satisfies individual rationality, strategy-proofness,
and the extensions of $(\Phi_i)_{i\in \mathcal{I}}$.\footnote{Some of the steps
in this proof are similar to the proof of Theorem 2 in \cite{sonmez_yenmez_india2} for the characterization of DA with the 2-step horizontal meritorious choice rule.}

\begin{lemma}
$DA$ based on $C$ satisfies individual rationality.
\end{lemma}

\begin{proof}
Let $X$ be the outcome of $DA$ based on $C$ for preference
profile $R=(R_a)_{a\in \mathcal{A}}$. In $DA$ based on $C$, agents only propose with acceptable
contracts. Therefore, for each agent $a\in \mathcal{A}$, $X_a \mathrel{R_a} \emptyset$, which means that $DA$ based on $C$ satisfies individual rationality.
\end{proof}

\begin{lemma}\label{lem:strategy}
$DA$ based on $C$ is strategy-proof.
\end{lemma}

\begin{proof}
Since institution choice rules are path independent,
the cumulative offers rule defined in \cite{hatfi04} is equivalent to
$DA$ based on $C$. Furthermore, since institution choice rules satisfy size monotonicity
in addition to path independence, the cumulative offers rule, which is
equal to $DA$ based on $C$, is strategy-proof \citep{hatfi04}.
\end{proof}

The following lemma establishes when extensions of punctual
choice axioms are satisfied.

\begin{lemma}\label{lem:punctual}
Consider an institution $i\in \mathcal{I}$, a matching problem
$R$, and a matching $X\subseteq \mathcal{X}$
such that $C_i(D_i(X))=X_i$. If $C_i$ satisfies a punctual choice axiom
$\phi_i$, then $X$ satisfies the matching axiom $\overline{\phi}_i$ at $R$.
\end{lemma}

\begin{proof}
Let $Y_i=D_i(X)$. Since choice rule $C_i$ satisfies the punctual axiom $\phi_i$,
we have
\[C_i(Y_i) \in \phi_i(Y_i).\]
Since $C_i(Y_i)=C_i(D_i(X))=X_i$ by assumption and $Y_i=D_i(X)$ by construction, the equation displayed above is equivalent to
\[X_i \in \phi_i(D_i(X)).\]
Therefore, $X$ satisfies the matching axiom $\overline{\phi}_i$.
\end{proof}

\begin{lemma}
$DA$ based on $C$ satisfies the extensions of $(\Phi_i)_{i\in \mathcal{I}}$.
\end{lemma}

\begin{proof}
Fix an institution
$i\in \mathcal{I}$ and a punctual choice axiom $\phi_i \in \Phi_i$.
Since $C_i$ is path independent, we have that $C_i(D_i(X))=X_i$.
Therefore, Lemma \ref{lem:punctual} implies that $X$ satisfies
$\overline{\phi}_i$. Therefore, $DA$ based on $C$ satisfies $\overline{\phi}_i$.
\end{proof}

\medskip

\textit{Only if part:} To finish the proof, using the following
lemmas, we next show that if a matching rule satisfies
individual rationality, strategy-proofness, and the extensions of
$(\Phi_i)_{i\in \mathcal{I}}$, then it has to be $DA$ based on $C$.

\begin{lemma}\label{lem:choice}
Let $X$ be the outcome of a matching rule that satisfies
the extensions of $(\Phi_i)_{i\in \mathcal{I}}$. Then $C_i(D_i(X))=X_i$.
\end{lemma}

\begin{proof}
Fix an institution $i \in \mathcal{I}$. Define the following choice rule $\tilde{C}_i$. For each $Y_i \subseteq \mathcal{X}_i$,
\[
     \tilde C_i(Y_i) = \begin{cases}
        C_i(Y_i), & \text{if } Y_i \neq D_i(X)\\
        X_i, & \text{if } Y_i = D_i(X).\\
        \end{cases}
  \]
We show that each punctual axiom $\phi_i \in \Phi_i$ is satisfied by $\tilde{C}_i$.
Therefore, we need to show that, for each $Y_i\subseteq \mathcal{X}_i$, $\tilde{C}_i(Y_i)\in \phi_i(Y_i)$. We consider two cases depending
on the value of $Y_i$.

\smallskip
\noindent
\emph{Case 1} ($Y_i\neq D_i(X)$): In this case, by construction of $\tilde{C}_i$,
we have $\tilde C_i(Y_i)=C_i(Y_i)$. Since $C_i$ satisfies $\phi_i$, we have
\[C_i(Y_i) \in \phi_i(Y_i).\]
Since $\tilde C_i(Y_i)=C_i(Y_i)$, the equation displayed above is equivalent to
\[\tilde{C}_i(Y_i)\in \phi_i(Y_i).\]

\smallskip
\noindent
\emph{Case 2} ($Y_i = D_i(X)$): Since the matching rule satisfies
the extension of $\phi_i$, we have
\[X_i \in \phi_i(D_i(x)).\]
Since $\tilde{C}_i(Y_i)=X_i$ by construction and $Y_i=D_i(X)$,
the equation displayed above is the same as
\[\tilde{C}_i(Y_i) \in \phi_i(Y_i).\]
\smallskip

In both cases we have shown that $\tilde{C}_i(Y_i)\in \phi_i(Y_i)$, which means that $\tilde{C}_i$ satisfies the punctual axiom $\phi_i$.
Since $\Phi_i$ characterizes $C_i$, we get $\tilde{C}_i=C_i$. Therefore, $X_i=\tilde{C}_i(D_i(X))=C_i(D_i(X))$, so $C_i(D_i(X))=X_i$.
\end{proof}

We need the following definition of stability for the next lemma \citep{gale62}.
A matching $X$ is \emph{stable} with respect to $(C_i)_{i\in \mathcal{I}}$
if
    \begin{itemize}
        \item (\textit{individual rationality\/}) for each $a \in \mathcal{A}$, \;  $X_a \succeq_a \emptyset$,
        \item (\textit{institution rationality\/}) for each  $i\in \mathcal{I}$,  \; $C_i(X_i)=X_i$, and
        \item (\textit{no blocking pairs\/}) there exist no $i \in \mathcal{I}$,
        $a\in \mathcal{A}$, and $x\in \mathcal{X}_a \cap \mathcal{X}_i$ such that
        \[x\in C_i(X_i \cup \{x\}) \mbox{ and } x \mathrel{\succ_a} X_a.\]
    \end{itemize}

\begin{lemma}\label{lem:stability}
If a matching rule satisfies individual rationality and the
extensions of $(\Phi_i)_{i\in \mathcal{I}}$, then it is stable
with respect to $(C_i)_{i\in \mathcal{I}}$.
\end{lemma}

\begin{proof}
Fix a matching problem $R$. Let $X$ be the matching produced for $R$ by the
given matching rule. Then by Lemma \ref{lem:choice}, $C_i(D_i(X))=X_i$. Therefore,
by path independence, $C_i(X_i)=X_i$, so institution rationality is satisfied.
To show that there are no blocking pairs, consider
$a\in \mathcal{A}$ and $x\in \mathcal{X}_a \cap \mathcal{X}_i$ such that
$x \mathrel{\succ_a} X_a$. Then $x\in D_i(X)\setminus X_i$. Since $C_i(D_i(X))=X_i$, path independence implies that $C_i(X_i \cup \{x\})=X_i$. Therefore,
$x \notin C_i(X_i \cup \{x\})$,
which implies that there are no blocking pairs.
Since the matching rule also satisfies individual rationality,
$X$ is stable with respect to $(C_i)_{i\in \mathcal{I}}$.
\end{proof}

\begin{lemma}\label{lem:da}
If a matching rule satisfies strategy-proofness and stability with respect to
$(C_i)_{i\in \mathcal{I}}$, then it is $DA$ based on $C$.
\end{lemma}

\begin{proof}
By Lemma \ref{lem:strategy}, $DA$ based on $C$ is strategy-proof. Furthermore, since
institution choice rules are path independent, the cumulative offers rule
defined in \cite{hatfi04} is stable and equivalent to $DA$ based on $C$. Hence,
$DA$ based on $C$ is stable with respect to $(C_i)_{i\in \mathcal{I}}$.
By Theorem 1 in \cite{HK17}, there can be at most
one stable and strategy-proof matching rule when choice rules
satisfy the \emph{irrelevance of rejected contracts} condition, a choice
rule axiom weaker than path independence. Therefore, the only matching rule
that satisfies strategy-proofness and stability with respect to
$(C_i)_{i\in \mathcal{I}}$ is $DA$ based on $C$.
\end{proof}

By Lemma \ref{lem:stability}, if a matching rule satisfies individual rationality
and the extensions of $(\Phi_i)_{i\in \mathcal{I}}$, then it is stable
with respect to $(C_i)_{i\in \mathcal{I}}$. By Lemma \ref{lem:da}, if it is
stable with respect to $(C_i)_{i\in \mathcal{I}}$ and strategy-proof, then it
has to be $DA$ based on $C$. Therefore, we conclude that, if a matching rule satisfies
individual rationality, strategy-proofness, and the extensions of $(\Phi_i)_{i\in \mathcal{I}}$, then it has to be $DA$ based on $C$, completing the proof.

\bigskip

\section{Proof of Proposition \ref{proposition_GE}}
\textit{If part:} We first show that $C_i^{ge}$ satisfies the four axioms stated in
the proposition. $C_i^{ge}$ satisfies guaranteed enrollment for returning students by Step 0. $C_i^{ge}$ satisfies non-wastefulness by Step 2. We show that $C_i^{ge}$ also satisfies the other two axioms.

Let $X_i^{ge}\equiv X_i\cap\mathcal{X}_i^{ge}$. Given $X_i^{ge}\subseteq\mathcal{X}_i^{ge}$, define the following function on $\mathcal{X}_i\setminus\mathcal{X}_i^{ge}$: for each $Y\subseteq\mathcal{X}_i\setminus\mathcal{X}_i^{ge}$,
\[r_i(Y|X_i^{ge})\equiv r_i(Y\cup X_i^{ge})-r_i(X_i^{ge})\]
Based on $r_i(\cdot|X_i^{ge})$, define
\[\mathcal{F}_i(X_i^{ge})\equiv \{Y\subseteq\mathcal{X}_i\setminus\mathcal{X}_i^{ge}: r_i(Y|X_i^{ge})=|Y|\}.\]
$(\mathcal{X}_i\setminus\mathcal{X}_i^{ge},\mathcal{F}_i(X_i^{ge}))$ is called the \emph{minor} of $(\mathcal{X}_i,\mathcal{F}_i)$ and is a matroid: see e.g., \cite{oxley}.
In addition, define
\[\mathcal{F}_i(X_i^{ge},q)\equiv \{Y\subseteq\mathcal{X}_i\setminus\mathcal{X}_i^{ge}: r_i(Y|X_i^{ge})=|Y|, |Y|\le q-r_i(X_i^{ge})\}.\]
$(\mathcal{X}_i\setminus\mathcal{X}_i^{ge},\mathcal{F}_i(X_i^{ge},q))$ is called the \emph{truncation} of $(\mathcal{X}_i\setminus\mathcal{X}_i^{ge},\mathcal{F}_i(X_i^{ge}))$ and is a matroid: see e.g., \cite{oxley}.
Let $C^m_i(\cdot|X_i^{ge})$ be the non-wasteful matroid choice rule (defined in Section \ref{sec:nw-matroid}) on $(\mathcal{X}_i\setminus\mathcal{X}_i^{ge},\mathcal{F}_i(X_i^{ge},q))$ where the capacity is $q-r_i(X_i^{ge})$.
Then, $C_i^{ge}(X)=X_i^{ge}\cup C^m_i(X\setminus X_i^{ge}|X_i^{ge})$ for each $X_i\subseteq\mathcal{X}_i$.

\begin{lemma}
$C_i^{ge}$ satisfies maximal utilization of reservations.
\end{lemma}
	
\begin{proof}
Since $C_i^{ge}$ satisfies non-wastefulness, for each $X_i\subseteq\mathcal{X}_i$, $X_i\setminus C_i^{ge}(X)\neq\emptyset$ imply $|C_i^{ge}(X)|=q$. Suppose, towards a contradiction, that there exist $X\subseteq \mathcal{X}$, $x\in X_i\setminus C_i^{ge}(X)$, and $y\in C_i^{ge}(X)\setminus \mathcal{X}^{ge}_i$ such that $|C_i^{ge}(X)|=q$ and
	\[
	r_i((C_i^{ge}(X)\setminus\{y\})\cup\{x\})> r_i(C_i^{ge}(X)).
	\]

Since $C_i^{ge}$ satisfies guaranteed enrollment for returning students, $x\not\in\mathcal{X}^{ge}_i$. Thus, $(C_i^{ge}(X)\setminus\{y\})\cup\{x\})\cap\mathcal{X}_i^{ge}=X_i^{ge}$. Let $X^*_i=C_i^{ge}(X)\setminus X_i^{ge}$ and $Y_i^*=(C_i^{ge}(X)\setminus\{y\}\cup\{x\})\setminus X_i^{ge}$. Then, we have
\begin{align*}
    r_i(C_i^{ge}(X)\setminus\{y\})\cup\{x\})&=r_i(Y_i^*|X_i^{ge})+r_i(X_i^{ge})\\
    &>r_i(X^*_i|X_i^{ge})+r_i(X_i^{ge})\\
    &=r_i(C_i^{ge}(X)).
\end{align*}
Thus, $r_i(Y_i^*|X_i^{ge})>r_i(X^*_i|X_i^{ge})$. However, $X^*_i=C^m_i(X\setminus X_i^{ge}|X_i^{ge})$, which is a contradiction since $C^m_i(\cdot|X_i^{ge})$ satisfies matroidal objectives by Proposition \ref{proposition_matroid}.


\end{proof}

\begin{lemma}
$C_i^{ge}$ satisfies no justified envy under reserves.
\end{lemma}

\begin{proof}
 Suppose, towards a contradiction, that there exist $X\subseteq \mathcal{X}$, $x\in C_i^{ge}(X)\setminus \mathcal{X}^{ge}_i$, and $y\in X_i\setminus C_i^{ge}(X)$ such that
	\[
	y\succ_i x \text{ and } r_i((C_i^{ge}(X)\setminus\{x\})\cup\{y\})\ge r_i(C_i^{ge}(X)).
	\]
	Let $X^*_i=C_i^{ge}(X)\setminus X_i^{ge}$ and
	$Y_i^*=(C_i^{ge}(X)\setminus\{x\}\cup\{y\})\setminus X_i^{ge}$. Then, we have
	\begin{align*}
        r_i(C_i^{ge}(X)\setminus\{x\})\cup\{y\})&=r_i(Y_i^*|X_i^{ge})+r_i(X_i^{ge})\\
        &\ge r_i(X^*_i|X_i^{ge})+r_i(X_i^{ge})\\
        &=r_i(C_i^{ge}(X)).
    \end{align*}
    Thus, $r_i(Y_i^*|X_i^{ge})\ge r_i(X^*_i|X_i^{ge})$. However, $X^*_i=C^m_i(X\setminus X_i^{ge}|X_i^{ge})$ and $y\succ x$, which is a contradiction since $C^m_i(\cdot|X_i^{ge})$ satisfies no justified envy under rank by Proposition \ref{proposition_matroid}.
\end{proof}

\textit{Only if part:} Let $C_i$ be an arbitrary choice rule satisfying  the four axioms stated in the proposition and let $X\subseteq \mathcal{X}$ be an arbitrary set of contracts
associated with $i\in \mathcal{I}$. We will show that each contract which is chosen in
the $C^{ge}_i$ procedure also belongs to $C_i(A)$, which will establish that $C_i(X)=C^{ge}_{i}(X)$ since both $C_i$ and $C^{ge}_i$ satisfy non-wastefulness.

All contracts which are chosen in Step $0$ of the $C_i^{ge}$ procedure must also belong
to $C_i(X)$ since $C_i$ satisfies guaranteed enrollment for returning students.
	
Suppose that there is a contract which is chosen before Step $2$ of the $C_i^{ge}$ procedure but does not belong to $C_i(X)$. Let Step $1.s$ be the earliest such step and let $x^s$ be the contract identified at that step. Note that $x^s$ is the highest priority contract in $X_i \setminus X^{s-1}_i$
	such that
	$r_i(X^{s-1}_i \cup \{x^s\})=r_i(X^{s-1}_i)+1$.  By non-wastefulness, $|C_i(X)|=q_i$ and $C_i(X)\setminus C^{ge}_i(X)\neq \emptyset$.
	
	\textit{Case 1:} For each $x\in C_i(X)\setminus C^{ge}_i(X)$, $x\succ_i x^s$.
	
	Since $x^s$ is the highest priority contract in $X_i \setminus X^{s-1}_i$ which can increase reserve utilization, and since all contracts which are chosen before Step $1.s$ in the $C^{ge}_i$ procedure also belong to $C_i(X)$, no contract in $C_i(X)\setminus C^{ge}_i(X)$ can increase reserve utilization once all contracts in $X^s_i$ are chosen. Take any $y\in C_i(X)\setminus C^{ge}_i(X)$. Note that \[
	r_i((C_i(X)\setminus\{y\})\cup\{x^s\})> r_i(C_i(X)),
	\]
	contradicting that $C_i$ satisfies maximal utilization of reservations.
	
	\textit{Case 2:} There exists $y\in C_i(X)\setminus C^{ge}_i(X)$ such that $x^s\succ_i y$.
	
	We begin with a lemma.
	\begin{lemma}\label{lem:submodular}
	For all $X\subseteq X'\subseteq\mathcal{X}$ and $x\in\mathcal{X}$,
	\[
	r_i(X'\cup\{x\})-r_i(X')\le r_i(X\cup\{x\})-r_i(X).
	\]
	\begin{proof}
	By R3 and $X\subseteq X'$,
	\begin{align*}
	   r_i((X\cup\{x\})\cup X')+r_i((X\cup\{x\})\cap X')&=r_i(X'\cup \{x\})+r_i(X) \\
	   &\le r_i(X\cup\{x\})+r_i(X').
	\end{align*}
	\end{proof}
	
	\end{lemma}
	Since $C_i$ satisfies no justified envy under reserves,
	\[
	r_i((C_i(X)\setminus\{y\})\cup\{x^s\})< r_i(C_i(X)).
	\]
	By R2, $r_i(C_i(X))\le r_i(C_i(X)\cup\{x^s\})$ and thus
	\[
	r_i((C_i(X)\setminus\{y\})\cup\{x^s\})< r_i(C_i(X)\cup\{x^s\}).
	\]
	By Lemma \ref{lem:submodular}, $r_i(X^{s}_i \cup \{y\})=r_i(X^{s}_i)+1$ and $r_i(X^{s-1}_i \cup \{y\})=r_i(X^{s-1}_i)+1$. Since $y\in X\setminus C^{ge}_i(X)$, $|X^{s}|=q$ and $C^{ge}_i(X)=X^s_i$. Note $X^{s-1}_i\subseteq C_i(X)$, which implies $C_i(X)=X^{s-1}_i\cup \{y\}$. These facts together imply
	\begin{align*}
    	r_i((C_i(X)\setminus\{y\})\cup\{x^s\})&= r_i(X^{s-1}_i \cup \{x^s\}) \\
    	&=r(C^{ge}_i(X)) \\
    	&= r_i(X^{s-1}_i \cup \{y\}) \\
	    &=r(C_i(X)),
	\end{align*}
    contradicting that $C_i$ satisfies no justified envy under reserves.
	
	Finally, suppose that all contracts chosen in Step $1$ of the $C^{ge}_i$ procedure also belong to $C_i(X)$ and there is a contract which is chosen in Step $2$ of the $C^{ge}_i$ procedure, say $x$, such that $x\notin C_i(X)$. By non-wastefulness, there is another contract $y\in C_i(X)\setminus C^{ge}_i(X)$, and by definition of the $C^{ge}_i$ procedure, $x\succ_i y$. Note that \[
	r_i((C_i(X)\setminus\{y\})\cup\{x\})= r_i(C_i(X)),
	\]
	contradicting that $C_i$ satisfies no justified envy under reserves.
	
	 Next, we show that $C^{ge}_i$ satisfies path independence and size monotonicity. $C^{ge}_i$ satisfies size monotonicity since $C^{ge}_i$ satisfies non-wastefulness.
	We first prove three lemmas for path independence.
		\begin{lemma}
		\label{addition}
		Let $X\subseteq \mathcal{X}$ and $x\in \mathcal{X}\setminus X$. If $Y$ is a base for $X$ and $Y\cup \{x\}\in \mathcal{F}_i$, then $Y\cup\{x\}$ is a base for $X\cup\{x\}$.
		\end{lemma}
	
	\begin{proof}
		Suppose not. Then, there exists $y\in X\cup \{x\}$ such that $Y\cup\{x,y\} \in \mathcal{F}_i$. Note that $y\in X$. By I2,  $Y\cup\{y\} \in \mathcal{F}_i$, which contradicts that $Y$ is a base for $X$.
		\end{proof}

	\begin{lemma}
		\label{cardinality}
		Let $X\subseteq \mathcal{X}$ and $x\in \mathcal{X}\setminus X$. Let $Y$ be a base for $X$ and $Y'$ be a base for $X\cup \{x\}$. If $Y$ is not a base for $X\cup \{x\}$, then $|Y'|=|Y|+1$.
		\end{lemma}

	\begin{proof}
		If $Y$ is a base for $X$ but not a base for $X\cup\{x\}$, then it must be that $Y\cup \{x\}\in
		\mathcal{F}_i$. By Lemma \ref{addition}, $Y\cup\{x\}$ is a base for $X\cup \{x\}$. Since all bases of $X\cup \{x\}$ have the same cardinality, $|Y'|=|Y\cup \{x\}|=|Y|+1$.
		\end{proof}
	
	
		
		
		\begin{lemma}
		\label{rank_to_base}
		 Let $X\subseteq \mathcal{X}$ and $x\in \mathcal{X}\setminus X$. Suppose that $r_i(X\cup \{x\})=r_i(X)+1$ and $Y$ is a base for $X$. Then, $Y\cup \{x\}$ is a base for $X\cup \{x\}$.
		\end{lemma}
		
		\begin{proof}
			Since $Y$ is a base for $X$, $r(X)=|Y|$. Since $r(X\cup \{x\})=r(X)+1$, there exists a base $Y'$ for $X\cup \{x\}$ such that $|Y'|=|Y|+1$. By I3, there exists $y\in Y'\setminus Y$ such that $Y\cup \{y\}\in \mathcal{F}_i$. If $y\neq x$, then $Y\cup \{y\}\subseteq X$, contradicting that $Y$ is a base for $A$. Therefore, $y=x$ and $Y\cup \{x\}\in \mathcal{F}_i$. If $Y\cup \{x\}$ is not a base for $X\cup \{x\}$, then there exists $z\in (X\cup \{x\})\setminus (Y\cup \{x\})$ such that $Y\cup \{x,z\}\in \mathcal{F}_i$. By I2, $Y\cup \{z\}\in \mathcal{F}_i$, contradicting that $Y$ is a base for $X$. Hence, $Y\cup \{x\}$ is a base for $X\cup \{x\}$\end{proof}
		
We want to show that $C_i^{ge}$ satisfies \emph{substitutability}: i.e., for each $X\subseteq \mathcal{X}$ and $x\in X$, $C_i^{ge}(X)\setminus \{x\}\subseteq C_i^{ge}(X\setminus \{x\})$. Substitutability and size monotonicity together imply path  independence. Let $X'=X\setminus\{x\}$. By construction, all contracts in $(X\cap \mathcal{X}^{ge}_i)\setminus \{x\}$ are chosen both in $C_i^{ge}(X)$ and $C_i^{ge}(X')$.
	
	We next show that any contract which is identified in Step 1 of the $C_i^{ge}$ procedure at $X$, and is different than $x$, is chosen also in the $C_i^{ge}$ procedure at $X'$. Towards a contradiction, suppose $y\in X'$ is such that $y$ is identified in Step $1.s$ of the $C_i^{ge}$ procedure at $X$, but not identified in any step of the $C_i^{ge}$ procedure at $X'$. Without loss of generality, assume that Step $1.s$ is the earliest such step of the $C_i^{ge}$ procedure at $X$. Note that $X^{s-1}_i\setminus \{x\}\subseteq C_i^{ge}(X)\cap C_i^{ge}(X')$.
	
	Since $y$ is identified in Step $1.s$ of the $C_i^{ge}$ procedure at $X$, by construction, $r_i(X^{s-1}_i\cup \{y\})=r_i(X^{s-1}_i)+1$. Let $Y$ be a base for $X^{s-1}_i$. By Lemma \ref{rank_to_base}, $Y\cup\{y\}$ is a base for $X^{s-1}_i\cup\{y\}$. Since $y$ is not identified in Step $1.s$ of the $C_i^{ge}$ procedure at $X'$, there exists $z\succ_i y$ such that $z$ is identified in Step $1.s$ of the $C_i^{ge}$ procedure at $X'$. Note that $x\in X^{s-1}_i$ since otherwise, instead of $y$, $z$ would be identified in Step $1.s$ of the $C_i^{ge}$ procedure at $X$.
	
	Since $z$ is identified in Step $1.s$ of the $C_i^{ge}$ procedure at $X'$, by construction, $r_i((X^{s-1}_i\setminus \{x\})\cup\{z\})=r_i(X^{s-1}_i\setminus \{x\})+1$. Let $Y'$ be a base for $X^{s-1}_i\setminus \{x\}$. By Lemma  \ref{rank_to_base}, $Y'\cup\{z\}$ is a base for $(X^{s-1}_i\setminus \{x\})\cup\{z\}$.
	
	We claim that $Y'$ is not a base for $X^{s-1}_i$. Suppose not, i.e., suppose that $Y'$ is a base for $X^{s-1}_i$. Since $Y'\cup\{z\}$ is a base for $(X^{s-1}_i\setminus \{x\})\cup\{z\}$, $Y'\cup\{z\}\in \mathcal{F}_i$. Then, $r_i(X^{s-1}_i\cup \{z\})=r_i(X^{s-1}_i)+1$, contradicting that $z$ is not identified in Step $1.s$ of the $C_i^{ge}$ procedure at $X$. Hence, $Y'$ is not a base for $X^{s-1}_i$.
	
	Since $Y'$ is a base for $X^{s-1}_i\setminus \{x\}$ but not a base for $X^{s-1}_i$, while $Y$ is a base for $X^{s-1}_i$, by Lemma \ref{cardinality}, $|Y|=|Y'|+1$, which implies that $|Y\cup\{y\}|=|Y'\cup\{z\}|+1$. By I3, there exists $w\in (Y\cup\{y\})\setminus (Y'\cup \{z\})$ such that $Y'\cup\{z,w\}\in \mathcal{F}_i$.
	
\textit{Case 1:} $w\in Y\setminus \{x\}$. Note that $Y\setminus \{x\}\subseteq X^{s-1}_i\setminus \{x\}$. Then, $Y'\cup \{z,w\}\in \mathcal{F}_i$ contradicts that $Y'\cup\{z\}$ is a base for $(X^{s-1}_i\setminus \{x\})\cup \{z\}$.
	
\textit{Case 2:} $w=x$. Since $Y'\cup\{z\}$ is a base for $(X^{s-1}_i\setminus \{x\})\cup\{z\}$ and $Y'\cup\{z,x\}\in \mathcal{F}_i$, then by Lemma \ref{addition}, $Y'\cup\{z,x\}$ is a base for $X^{s-1}_i\cup\{z\}$. Note that $|Y'\cup\{z,x\}|=|Y\cup \{y\}|$. Now, since $r_i(X^{s-1}_i\cup \{y\})=r_i(X^{s-1}_i)+1$, also  $r_i(X^{s-1}_i\cup \{z\})=r_i(X^{s-1}_i)+1$, contradicting that $z$ is not identified in the $C_i^{ge}$ procedure at $X$.
	
\textit{Case 3:} $w=y$. Since $Y'\cup\{z\}$ is a base for $(X^{s-1}_i\setminus \{x\})\cup\{z\}$ and $Y'\cup\{y,z\}\in \mathcal{F}_i$, we have $r_i((X^{s-1}_i\setminus \{x\})\cup\{y,z\})=r_i((X^{s-1}_i\setminus \{x\})\cup\{z\})+1$. Because $y\notin C_i^{ge}(X')$, there exists $v\succ_i y$ such that $v$ is identified in Step 1.(s+1) of the $C_i^{ge}$ procedure at $X'$. Then, by Lemma \ref{rank_to_base}, $Y'\cup \{z,v\}\in \mathcal{F}_i$ is a base for $X^{s-1}_i\cup \{z,v\}$.  Since $|Y|=|Y'|+1$, $|Y'\cup \{z,v\}|>|Y|$. By I3, there exists $w'\in \{z,v\}$ such that $Y\cup\{w'\}\in \mathcal{F}_i$, which contradicts that neither $z$ nor $v$ is identified in the $C_i^{ge}$ procedure at $X$.

Finally, we show that any contract which is chosen in Step 2 of the $C_i^{ge}$ procedure at $X$ is also chosen in $C_i^{ge}(X')$. Note that it will suffice to show that the total number of contracts chosen (including those in $\mathcal{X}^{ge}_i$) by the end of Step 1 of the $C_i^{ge}$ procedure at $X$ is greater than or equal to the total number of contracts chosen by the end of Step 1 of the $C_i^{ge}$ procedure at $X'$. Since we have shown that all contracts which are identified in Step 1 of the $C_i^{ge}$ procedure at $X$ are still chosen in $C_i^{ge}(X')$, it will suffice to show that there cannot be any two contracts in $C_i^{ge}(X')\setminus C_i^{ge}(X)$ which are identified in Step 1 of the $C_i^{ge}$ procedure at $X'$. Towards a contradiction, suppose that $z,v\in C_i^{ge}(X')\setminus C_i^{ge}(X)$ are identified in Step 1 of the $C_i^{ge}$ procedure at $X'$. Without loss of generality, suppose that $v$ is identified at a later step than $z$, say Step 1.t. Let $X''$ be the set of all contracts which are identified after Step 1.s and until, and including, Step 1.t. Note that $z,v\in X''$. By repeated application of Lemma \ref{rank_to_base}, $Y'\cup X''$ is a base for $X^{s-1}_i\cup X''$. By I2, $Y'\cup \{z,v\}\in \mathcal{F}_i$. Since $|Y|=|Y'|+1$, $|Y'\cup \{z,v\}|>|Y|$. By I3, there exists $w'\in \{z,v\}$ such that $Y\cup\{w'\}\in \mathcal{F}_i$, which contradicts that neither $z$ nor $v$ is identified in the $C_i^{ge}$ procedure at $X$.

\bigskip

\section{Proof of Proposition \ref{proposition_matroid}}
Let $Y^*(X)$ be the set of contracts chosen in Step 1 of the $C_i^m$ procedure at $X$. By construction, $Y^*(X)$ is a base for $X$. We will invoke the following known facts.



\begin{lemma}[\citeauthor{gale1968optimal},\citeyear{gale1968optimal}]\label{lem:gale}
For any base $Y$ of $X$, the highest priority contract in $Y^*(X)$ has weakly higher priority than the highest priority contract in $Y$, the second highest priority contract in $Y^*(X)$ has weakly higher priority than the second highest priority contract in $Y$, and so on.
\end{lemma}

Let us introduce the following standard base axiom of a matroid: see e.g., \cite{oxley}.
\begin{itemize}
    \item[(B1)] If $Y$ and $Y'$ are both bases of $X$ and $x\in Y\setminus Y'$, then there exists $y\in Y'\setminus Y$ such that $(Y\setminus \{x\})\cup \{y\}$ is also a base of $X$.
\end{itemize}
\medskip


\textit{If part:} Let $X\subseteq \mathcal{X}$. Since all bases of $X$ have the same cardinality, $Y^*(X)\subseteq C_i^{m}(X)$ has the same cardinality with all bases of $X$, and therefore $C_i^{m}$ satisfies matroidal objectives.

Suppose, towards a contradiction, that $C_i^{m}$ violates no justified envy. Then, there exist $X\subseteq \mathcal{X}$, $x\in C_i^{m}(X)$, and $y\in X\setminus C_i^{m}(X)$ such that $y\succ_i x$ and
$r_i((C_i^{m}(X)\setminus \{x\})\cup\{y\})\geq r_i(C_i^m(X)).$ Note that $x\in Y^*(X)$ because, by construction, any contract chosen in Step 2 of the $C_i^m$ procedure at $X$ has a higher priority than $y$.

Since $C_i^m$ satisfies matroidal objectives, $r_i(C_i^m(X))\geq r_i(X)$. Since $(C_i^m(X)\setminus \{x\})\cup\{y\}\subseteq X$, by R2, $r_i((C_i^m(X)\setminus \{x\})\cup\{y\})\leq r_i(X)$, which, together with the above observation, implies that $r_i((C_i^m(X)\setminus \{x\})\cup\{y\})= r_i(X)=r_i(C_i^m(X))$. Then, there exists a base $Y$ of $X$ such that $Y\subseteq (C_i^m(X)\setminus \{x\})\cup\{y\}$ and $|Y|=|Y^*(X)|$. Let $\{z\}=Y\setminus Y^*(X)$. By Lemma \ref{lem:gale}, $x\succ_i z$, and therefore $z\in C_i^m(X)\setminus Y^*(X)$, that is, $z$ is chosen in Step 2 of the $C_i^m$ procedure at $A$. By construction, $z\succ_i y$, which implies that $x\succ_i y$, a contradiction.

Note that $C_i^m$ satisfies non-wastefulness by Step 2.\medskip

\textit{Only if part:} Let $C_i$ be an arbitrary choice rule that satisfies the axioms. Let $X\subseteq \mathcal{X}$. We first show that $Y^*(X)\subseteq C_i(X)$. Since $C_i$ satisfies matroidal objectives, there exists a base $Y$ for $X$ such that $Y\subseteq C_i(X)$. Suppose that $Y\neq Y^*(X)$. Since all bases of $X$ have the same cardinality, there exists $x\in Y\setminus Y^*(X)$. By B1, there exists $y\in Y^*(X)\setminus Y$ such that $(Y\setminus \{x\})\cup \{y\}$ is also a base for $X$. By Lemma \ref{lem:gale}, $y\succ_i x$. If $y\notin C_i(X)$, this contradict that $C_i$ satisfies no justified envy because $r_i((C_i(X)\setminus \{x\})\cup \{y\})=r_i(C_i(X))$. So, $y\in C_i(X)$ and $(Y\setminus \{x\})\cup \{y\}\subseteq C_i(X)$. If $(Y\setminus \{x\})\cup \{y\}=Y^*(X)$, we are done. Otherwise, continuing similarly, we will establish that $Y^*(X)\subseteq C_i(X)$.

Now, suppose that $C_i(X)\setminus Y^*(X)\neq C_i^m(X)\setminus Y^*(X)$. Since both $C_i$ and $C_i^m$ satisfies non-wastefulness, there exist $x\in C_i(X)\setminus C_i^m(X)$ and $y\in C_i^m(X)\setminus C_i^(X)$. By construction of the $C_i^m$ procedure, $y\succ_i x$. Note that $r_i((C_i(X)\setminus \{x\})\cup\{y\})=r_i(C_i(X))$, which contradicts that $C_i$ satisfies no justified envy. Hence, $C_i(X)=C_i^m(X)$.
\bigskip


\section{Proof of Corollary \ref{cor:ge}}
To apply Theorem \ref{thm:main}, we need to show that the choice axioms (i) guaranteed enrollment for returning students, (ii) maximal utilization of reservations, and (iii) no justified envy under reserves are punctual and their extensions are matching axioms
with the same names. Note that in Section \ref{sec:punext} we show the
same property for non-wastefulness.

Let $X \in \mathcal{M}$ be a matching and $R=(R_a)_{a\in \mathcal{A}}$
be a matching problem. We first construct the correspondence for each
choice axiom listed above, hence, establishing that it is punctual,
then construct its extension, and finally show the equivalence to the matching
axiom with the same name.

First, guaranteed enrollment for returning students is a
punctual axiom because a choice rule satisfies it if and only if
it is a selection from the following correspondence:
\[\phi_i^{ge}(X_i)=\{Y_i \subseteq X_i : X_i \cap \mathcal{X}_i^{ge} \subseteq Y_i\}.\]
Therefore, the extension of this axiom can be stated as, for each $X_i \subseteq \mathcal{X}_i$,
\[X_i\in \phi_i^{ge}(D_i(X))=\{Y_i\subseteq D_i(X) : D_i(X) \cap
\mathcal{X}_i^{ge}  \subseteq Y_i\}.\]
Since $X_i\subseteq D_i(X)$, the extension is equivalent to
\[D_i(X)\cap \mathcal{X}_i^{ge} \subseteq X_i.\]
By the definition of $D_i(X)$, we can rewrite this as follows: For each $x\in \mathcal{X}_i \setminus X_i$
\[\alpha(x) \in \mathcal{A}_i \mbox{ implies } X_{\alpha(x)}\mathrel{R_{\alpha(x)}}x.\]
The contrapositive of the statement is the extension of
guaranteed enrollment for returning students.

Next, we consider maximal utilization of reservations. It is easy to see that
maximal utilization of reservations is a punctual axiom. The correspondence
for this axiom can be written as follows, for any $X_i\subseteq \mathcal{X}_i$,
\begin{flalign*}
\phi_i^{mr}(X_i) = & \{Y_i\subseteq X_i : |Y_i|<q_i, x \in X_i \setminus Y_i \Rightarrow r_i(Y_i\cup \{x\})=r_i(Y_i)\} \; \cup \\
     & \{Y_i\subseteq X_i : |Y_i|=q_i, x\in X_i \setminus Y_i, \mbox{ and }
     y\in Y_i \setminus \mathcal{X}_i^{ge} \Rightarrow     r_i((Y_i\setminus\{y\})\cup\{x\})\leq r_i(Y_i)\}.
\end{flalign*}
The extension of this axiom can be stated as, for each $X_i \subseteq \mathcal{X}_i$, $X_i\in \phi_i^{mr}(D_i(X))$ where $\phi_i^{mr}(D_i(X))$ is defined as
\begin{flalign*}
 & \{Y_i\subseteq X_i : |Y_i|<q_i, x \in D_i(X) \setminus Y_i \Rightarrow r_i(Y_i\cup \{x\})=r_i(Y_i)\} \; \cup \\
     & \{Y_i\subseteq X_i : |Y_i|=q_i, x\in D_i(X) \setminus Y_i, \mbox{ and }
     y\in Y_i \setminus \mathcal{X}_i^{ge} \Rightarrow     r_i((Y_i\setminus\{y\})\cup\{x\})\leq r_i(Y_i)\}.
\end{flalign*}
Since $X_i\subseteq D_i(X)$, we can rewrite this that for $x\in D_i(X)\setminus X_i$,
\begin{enumerate}
    \item [(i)] if $\abs{X_i}<q_i$, then $r_i(X_i\cup \{x\})=r_i(X_i)$, and
\item [(ii)] if $\abs{X_i}=q_i$, then for any
$y \in X_i \setminus \mathcal{X}_i^{ge}$, $r_i((X_i \setminus\{y\})\cup\{x\})\leq r_i(X_i).$
\end{enumerate}
Since $y\in D_i(X)\setminus X_i$, we have $y \mathrel{P_{\alpha(y)}} X_{\alpha(y)}$.
Together with these facts, we get the extension of maximal utilization of reservations.

Finally, we consider no justified envy under reserves. It can be easily
seen that no justified envy under reserves is also punctual:
\[\phi_i^{ne} (X_i) = \{Y_i\subseteq X_i: x\in Y_i \setminus \mathcal{X}_i^{ge},
y\in X_i \setminus Y_i, \mbox{ and } y \succ_i x  \Rightarrow  r_i((Y_i\setminus\{x\})\cup\{y\})<r_i(Y_i)\}.\] The extension of this axiom
to matchings is, for each $X_i\subseteq \mathcal{X}_i$, $X_i\in \phi_i^{ne}(D_i(X))$ where $\phi_i^{ne}$ as
\begin{align*}
    \{Y_i\subseteq D_i(X): x\in Y_i, \alpha(x)\notin\mathcal{A}_i, y\in D_i(X)\setminus Y_i, \mbox{ and } x \succ_i y \; \implies r_i((Y_i\setminus\{x\})\cup\{y\})<r_i(Y_i)\}.
\end{align*}
Since $X_i \subseteq D_i(X)$, we can rewrite this as
\[x\in X_i, \alpha(x)\notin\mathcal{A}_i, y\in D_i(X)\setminus X_i, \;  \mbox{and} \; x \succ_i y \; \implies r_i((X_i\setminus\{x\})\cup\{y\})<r_i(X_i).\]
Since $y\in D_i(X)\setminus X_i$, we have $y \mathrel{P_{\alpha(y)}} X_{\alpha(y)}$.
Together with these facts, we get the extension of no justified envy under reserves.
\bigskip

\section{Proof of Corollary \ref{cor:mat1}}
We follow the similar argument in the proof of Corollary \ref{cor:ge}. Given a matching $X \in \mathcal{M}$
and a matching problem $R=(R_a)_{a\in \mathcal{A}}$, we extend the punctual choice rule axioms to matchings as follows.
\medskip

Consider feasibility first. The extension of this axiom can be stated as, for each $X_i \subseteq \mathcal{X}_i$,
\[X_i\in \phi_i^{f}(D_i(X))=\{Y_i\subseteq D_i(X) : Y_i \in \mathcal{F}_i\}.\]
Since $X_i\subseteq D_i(X)$, the extension of feasibility is equivalent to
\[X_i \in \mathcal{F}_i.\]

Next, we consider rank maximality.
The extension of this axiom can be stated as, for each $X_i \subseteq \mathcal{X}_i$,
\[X_i\in \phi_i^{rm}(D_i(X))=\{Y_i\subseteq D_i(X) : y\in D_i(X)\setminus Y_i \; \implies \; r_i(Y_i\cup\{y\})= r_i(Y_i) \}.\]
Since $X_i\subseteq D_i(X)$, we can rewrite this as
\[y\in D_i(X)\setminus X_i \; \implies \; r_i(X_i\cup\{y\})= r_i(X_i).\]
Since $y\in D_i(X)\setminus X_i$, we have $y \mathrel{P_{\alpha(y)}} X_{\alpha(y)}$.
Together with these facts, we get the extension of rank maximality.

Finally, we consider no justified envy. The extension of this axiom
to matchings is, for each $X_i\subseteq \mathcal{X}_i$, $X_i\in \phi_i^{ne}(D_i(X))$ where $\phi_i^{ne}$ as
\begin{align*}
    \{Y_i\subseteq D_i(X) : x\in Y_i, y\in D_i(X)\setminus Y_i, \;  \mbox{and} \; x \succ_i y \; \implies r_i((Y_i\setminus\{x\})\cup\{y\})<r_i(Y_i)\}.
\end{align*}
Since $X_i \subseteq D_i(X)$, we can rewrite this as
\[x\in X_i, y\in D_i(X)\setminus X_i, \;  \mbox{and} \; x \succ_i y \; \implies r_i((X_i\setminus\{x\})\cup\{y\})<r_i(X_i).\]
Since $y\in D_i(X)\setminus X_i$, we have $y \mathrel{P_{\alpha(y)}} X_{\alpha(y)}$.
Together with these fact, we get the extension of no justified envy under rank.

\section{Strengthening Theorem \ref{thm:main}: An Impossibility}
\label{stronger_thm}

We show that the following statement, which would imply Theorem \ref{thm:main}, is not true.\smallskip

\textit{Alternative statement:} Suppose that, for each institution $i\in I$, the choice rule $C_i$ is characterized by path independence, size monotonicity, and a set of punctual axioms $\Phi_i$. Then a matching rule satisfies
individual rationality, strategy-proofness, and the extensions of
$(\Phi_i)_{i\in \mathcal{I}}$ if, and only if, it is $DA$ based on $C$.\smallskip

Consider the following example. Let $\mathcal{I}=\{i\}$, $\mathcal{A}=\{a,b\}$, and $\mathcal{X}=\{x,y\}$
where $\alpha(x)=a$ and $\alpha(y)=b$. 
	
Consider the following punctual choice axiom $\phi_i$ given by
$$\phi_i(\{x,y\})=\{\{x\}\},\quad \phi_i(\{x\})=\{\{x\},\emptyset\},\quad \phi_i(\{y\})=\{\emptyset\}$$
and choice rule $C_i$ given by
$$C_i(\{x,y\})=\{x\},\quad \C_i(\{x\})=\{x\},\quad C_i(\{y\})=\{\emptyset\}.$$
For each $X_i\subseteq \mathcal{X}_i$, $C_i(X_i)\in \phi_i(X_i)$, so $C_i$
satisfies $\phi_i$. On the other hand, $C_i$ is not characterized by $\phi_i$ since
for at the choice problem $\{x\}$, $\phi_i$ is multi-valued.
However, $C_i$ is characterized by path independence, size monotonicity, and $\phi_i$
because the only other choice rule that satisfies $\phi_i$ fails path independence.

Define the matching rule $\varphi$ as follows: for each problem $R=(R_a,R_b)$, if both agents find their contracts with $i$ acceptable, then $\varphi_a(R)=x$ and $\varphi_b(R)=\emptyset$; otherwise, $\varphi_a(R)=\varphi_b(R)=\emptyset$. First note that $\varphi\neq DA$ because, for example, for the problem $R$ where $a$ finds $x$ acceptable while $b$ does not find $y$ acceptable, $DA_a(R)=x$ and $\varphi_a(R)=\emptyset$. In addition, $\varphi$ satisfies individual rationality and strategy-proofness. It is easy to see that $\varphi$ satisfies the extension $\overline{\phi}_i$. If both agents find their contracts with $i$ acceptable, then $D_i(\varphi(R))=\{x,y\}$. Thus, $\phi_i(D_i(\varphi(R)))=\phi_i(\{x,y\})=\{\{x\}\}$, which imply $\varphi_i(R)=\{x\}\in\phi_i(D_i(\varphi(R)))$. Otherwise, $D_i(\varphi(R))=\{x\},\{y\}\mbox{ or }\os$. In all cases, $\emptyset\in\phi_i(D_i(\varphi(R)))$. Thus, $\varphi_i(R)=\emptyset\in\phi_i(D_i(\varphi(R)))$.

Hence, $DA$ based on $C$ is not characterized by individual rationality, strategy-proofness, and the extension $\overline{\phi}_i$.

\end{document}